\documentclass[10pt,twocolumn]{article}
\usepackage[hidelinks,bookmarks=false]{hyperref}
\usepackage[T1]{fontenc}
\usepackage{amsthm}
\usepackage{framed}
\usepackage{fullpage}
\usepackage{breakurl}
\usepackage[hang,flushmargin]{footmisc}
\usepackage{cite}
\usepackage{color}
\usepackage{paralist}
\usepackage{amssymb,amsmath}
\usepackage{epsfig}
\usepackage{url}
\usepackage{xspace}
\usepackage{subfig}
\usepackage[notcomma,notperiod,notquote,notcolon,notexcl,notquery,notscolon]{hanging}
\usepackage{mathptmx}

\DeclareMathAlphabet{\mathcal}{OMS}{cmsy}{m}{n}

\long\def\abbr#1#2{#2}     

\newtheorem{defn}{Definition}

\newcommand\blfootnote[1]{\begingroup
  \renewcommand\thefootnote{}\footnote{#1}\addtocounter{footnote}{-1}\endgroup
}

\newcommand{\E}{\textrm{E}}
\newcommand{\Z}{\mathbb{Z}}
\newcommand{\F}{\mathbb{F}}
\newcommand{\Seed}{\mathbb{S}}
\newcommand{\G}{\mathbb{G}}
\newcommand{\HH}{\mathcal{H}}
\newcommand{\rgets}{\mathrel{\mathpalette\rgetscmd\relax}}
\newcommand{\rgetscmd}{\ooalign{$\leftarrow$\cr
        \hidewidth\raisebox{1.2\height}{\scalebox{0.5}{\ \rm R}}\hidewidth\cr}}
\newcommand{\Adv}{\mathcal{A}}
\newcommand{\hbet}{\hat{\beta}}
\newcommand{\hsA}{\hat{\sigma}_A}
\newcommand{\hsB}{\hat{\sigma}_B}
\newcommand{\lbits}{\{0,1\}^\lambda}
\newcommand{\mA}{{\bf m}_A}
\newcommand{\hmA}{\hat{{\bf m}}_A}
\newcommand{\hcA}{\hat{c}_A}
\newcommand{\hcB}{\hat{c}_B}
\newcommand{\mB}{{\bf m}_B}
\newcommand{\hmB}{\hat{{\bf m}}_B}
\newcommand{\hvA}{\hat{{\bf v}}_A}
\newcommand{\vA}{{\bf v}_A}
\newcommand{\hvB}{\hat{{\bf v}}_B}
\newcommand{\vB}{{\bf v}_B}

\newcommand{\MM}{\mathcal{M}}
\newcommand{\state}{\textsf{state}}
\newcommand{\Gen}{\ensuremath{\mathsf{Gen}}\xspace}
\newcommand{\Eval}{\ensuremath{\mathsf{Eval}}\xspace}
\newcommand{\nicepara}[1]{\medskip\noindent\textbf{#1.}}

\newcommand{\name}{Riposte\xspace}
\newcommand{\Name}{Riposte\xspace}

\title{\Large \bf \Name: An Anonymous Messaging System\\Handling Millions of Users}
\author{Henry Corrigan-Gibbs, Dan Boneh, and David Mazi\`eres\\
Stanford University}
\date{\today}
\begin{document}

\maketitle

\begin{abstract}
This paper presents \name, a new system for anonymous
broadcast messaging.
\Name is the first such system, to our knowledge, that
simultaneously protects against traffic-analysis attacks,
prevents anonymous denial-of-service by malicious clients,
and scales to million-user anonymity sets.
To achieve these properties, \name makes novel use
of techniques used in systems for private information retrieval and
secure multi-party computation. 
For latency-tolerant workloads with many more readers than writers 
(e.g.~Twitter, Wikileaks), we demonstrate that
a three-server \name cluster can build an anonymity set of 2,895,216 users
in 32 hours.
\end{abstract}

\blfootnote{{\color{blue} \textbf{Note:}} 
This is the extended and corrected
version of a paper by the same name that appeared at the
{\em IEEE Symposium on Security and Privacy}
in May 2015. This version corrects an error in the 
${\sf AlmostEqual}$ protocol of Section~\ref{sec:disrupt:smc}
that could allow a malicious database server to de-anonymize
a client using an active attack.
We thank Elette Boyle and Yuval Ishai for pointing out this
error and for helpful discussions on how to correct it.
Since the complexity of the updated protocol is almost identical
to the original one, and since the first author's 
dissertation~\cite{thesis} describes and evaluates
a DPF-checking protocol that improves upon the one
presented here, the evaluation section (Section~\ref{sec:eval}) 
of this paper reflects the DPF-checking protocol from
the original version of this work.
}

\section{Introduction}

In a world of ubiquitous network 
surveillance~\cite{bennhold2014britain,gellman2013nsa,
gellman2014nsa,goel2014government,nakashima2014court}, 
prospective whistleblowers face a daunting task.
Consider, for example, a government employee who wants to anonymously 
leak evidence of waste, fraud, or incompetence to the public.
The whistleblower could email an investigative reporter directly, but {\em post hoc} 
analysis of email server logs could easily reveal the tipster's identity.
The whistleblower could contact a reporter via 
Tor~\cite{dingledine2004tor} or another
low-latency anonymizing proxy~\cite{freedman2002tarzan,leblond2013towards,mittal2009shadowwalker,reiter1998crowds}, 
but this would leave the leaker vulnerable to traffic-analysis attacks~\cite{bauer2007low,murdoch2005low,murdoch2007sampled}.
The whistleblower could instead use an anonymous messaging system that protects
against traffic analysis attacks~\cite{chaum1988dining,goel2003herbivore,wolinsky2012dissent}, 
but these systems typically only support
relatively small anonymity sets (tens of thousands of users, at most).
Protecting whistleblowers in the digital age requires 
anonymous messaging systems that provide strong security guarantees,
but that also scale to very large network sizes.

In this paper, we present a new system that attempts to make
traffic-analysis-resistant anonymous broadcast 
messaging practical at Internet scale.
Our system, called \name, allows a large number of clients 
to anonymously post messages to a
shared ``bulletin board,'' maintained by a small set of 
minimally trusted servers. 
(As few as three non-colluding servers are sufficient).
Whistleblowers could use \name as a platform for anonymously publishing Tweet-
or email-length messages and could combine it with standard
public-key encryption to build point-to-point private messaging channels.

While there is an extensive literature on anonymity 
systems~\cite{danezis2008survey,edman2009anonymity},
\name offers a combination of 
security and scalability properties unachievable with
current designs.
To the best of our knowledge, \name is the only anonymous
messaging system that simultaneously:
\begin{compactenum}
\item protects against traffic analysis attacks,
\item prevents malicious clients from anonymously executing 
      denial-of-service attacks, and 
\item scales to anonymity set sizes of 
      {\em millions} of users,
      for certain latency-tolerant applications.
\end{compactenum}
We achieve these three properties in \name 
by adapting three different techniques from the 
cryptography and privacy literature.
First, we defeat traffic-analysis attacks 
and protect against malicious servers by 
using a protocol, inspired by client/server 
DC-nets~\cite{chaum1988dining,wolinsky2012dissent},
in which every participating client sends a fixed-length
secret-shared message to the system's servers in every time epoch.
Second, we achieve efficient disruption resistance
by using a secure multi-party protocol 
to quickly detect and exclude malformed 
client requests~\cite{fagin1996comparing,goldreich1987play,yao1982protocols}.
Third, we achieve scalability
by leveraging a specific technique developed in the context of private information retrieval (PIR) to 
minimize the number of bits each client must upload to
each server in every time epoch.  The tool we use is called
a {\em distributed point function}~\cite{chor1997computationally,gilboa2014distributed}.
The novel synthesis of these techniques leads to a
system that is efficient (in terms of bandwidth and 
computation) and practical, even for large anonymity sets.

Our particular use 
of private information retrieval (PIR) protocols is unusual:
PIR systems~\cite{chor1998private}  allow a client to efficiently read a row from a database,
maintained collectively at a set of servers, without
revealing to the servers which row it is reading.
\Name achieves scalable anonymous messaging by running
a private information retrieval protocol {\em in reverse}:
with reverse PIR, a \name client can efficiently {\em write} into a database
maintained at the set of servers without revealing to 
the servers which row it has written~\cite{ostrovsky1997private}.

As we discuss later on, a large \name deployment 
could form the basis for an anonymous Twitter service.
Users would ``tweet'' by using \name to anonymously 
write into a database containing all clients' tweets for
a particular time period.
In addition, by having read-only users submit ``empty''
writes to the system, the effective anonymity set can
be much larger than the number of writers, with little impact
on system performance.

\medskip

Messaging in \name proceeds in regular {\em time epochs}
(e.g., each time epoch could be one hour long).
To post a message, the client generates 
a {\em write request}, cryptographically splits it into many
shares, and sends one share to each of the \name servers.
A coalition of servers smaller than a certain threshold cannot
learn anything about the client's message or write location given
its subset of the shares.

The \name servers collect write requests until the end of the time epoch,
at which time they publish the aggregation of the write requests they received
during the epoch.
From this information, anyone can recover the set of posts uploaded during
the epoch, but the system reveals no information about who posted which message.  
The identity of the entire set of clients who
posted during the interval is known, but no one can link a client to a post.
(Thus, each time epoch must be long enough to ensure that a large
number of honest clients are able to participate in each epoch.)

\begin{figure*}
  \centering
  \subfloat[A client submits one share of its write 
      request to each of the two database servers.
      If the database has length $L$, each share has
      length $O(\sqrt{L})$.]{\includegraphics[height=0.14\textwidth]{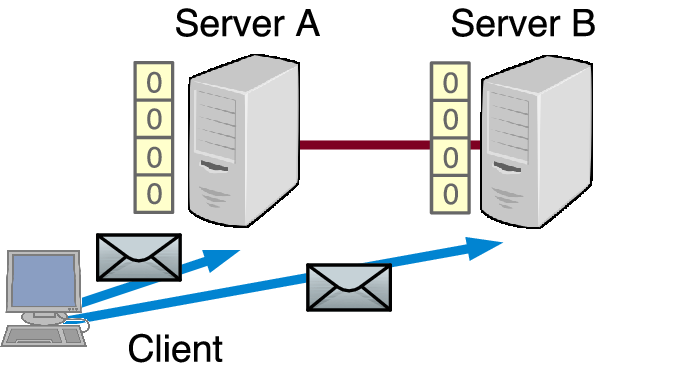}}
  \quad
  \subfloat[The database servers generate blinded ``audit request'' messages derived
      from their shares of the write request.]{\includegraphics[height=0.14\textwidth]{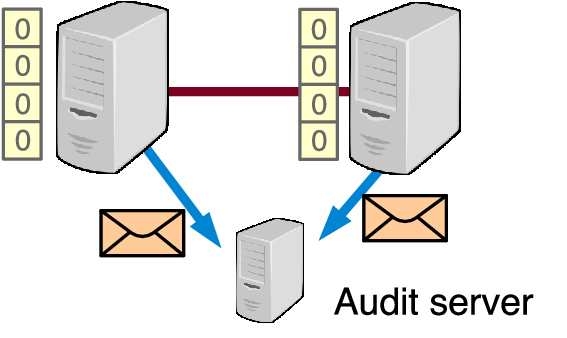}}
  \quad
  \subfloat[The audit server uses the audit request messages to validate the 
      client's request and returns an
      ``OK'' or ``Invalid'' bit to the database servers.]{\includegraphics[height=0.14\textwidth]{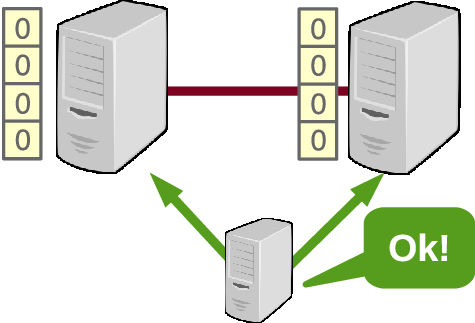}}
  \quad
  \subfloat[The servers apply the write request to their local database state.
    The XOR of the servers' states contains the clients message at the given row.]{\includegraphics[height=0.14\textwidth]{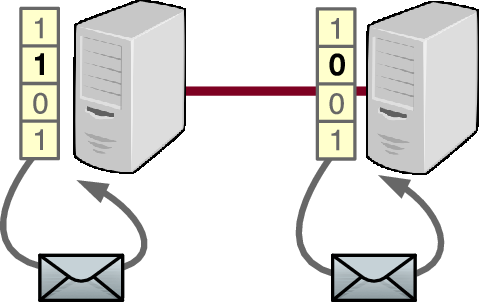}}

    \caption{The process of handling a single client write request. 
        The servers run this process once per client in each time epoch.}
    \label{fig:arch}
\end{figure*}

In this paper, we describe two \name variants, which
offer slightly different security properties.
The first variant scales to very large network
sizes (millions of clients) but
requires three servers such that no two of these servers collude.
The second variant is more computationally expensive, 
but provides security
even when all but one of the $s>1$ servers are malicious.
Both variants maintain their security properties
when network links are actively adversarial, when 
all but two of the clients are actively malicious, and
when the servers are actively malicious 
(subject to the non-collusion requirement above).

The three-server variant uses a computationally inexpensive
multi-party protocol to detect and exclude malformed client
requests. (Figure~\ref{fig:arch} depicts this
protocol at a high-level.)
The $s$-server variant uses client-produced
zero-knowledge proofs to guarantee the well-formedness of 
client requests.

Unlike Tor~\cite{dingledine2004tor} and
other low-latency anonymity systems~\cite{goel2003herbivore,
  hsiao2012lap,
  leblond2013towards,
  reiter1998crowds},
\name protects against active traffic analysis attacks
by a global network adversary.
Prior systems have offered traffic-analysis-resistance 
only at the cost of scalability:
\begin{compactitem}
\item
Mix-net-based systems~\cite{chaum1981untraceable} require
large zero-knowledge proofs of correctness to provide privacy in
the face of active attacks by malicious servers~\cite{adida2007shuffle,bayer2012efficient,furukawa2004efficient,groth2010verifiable,neff2001verifiable}.
\item
DC-nets-based systems require clients to transfer data {\em linear} in the size
of the anonymity set~\cite{chaum1988dining,wolinsky2012dissent}
and rely on expensive zero-knowledge proofs
to protect against malicious clients~\cite{corrigangibbs2013proactively,golle2004dining}.
\end{compactitem}
We discuss these systems and other prior work in Section~\ref{sec:rel}.

\nicepara{Experiments}
To demonstrate the practicality of \name for 
anonymous broadcast messaging 
(i.e., anonymous whistleblowing or microblogging), we
implemented and evaluated the complete three-server variant of the system.
When the servers maintain a database table large enough to fit
65,536 160-byte Tweets, the system can process 32.8 client write
requests per second.
In Section~\ref{sec:eval:million}, we discuss how to use a table of this
size as the basis for very large anonymity 
sets in read-heavy applications.
When using a larger 377 MB database table (over 2.3 million 160-byte Tweets),
a \name cluster can process 1.4 client write requests per second.

Writing into a 377 MB table requires
each client to upload less than 1 MB of data to the servers.
In contrast, a two-server DC-net-based system would 
require each client to upload more than 750 MB of data.
More generally, to process a \name client request for a table of size $L$,
clients and servers perform only $O(\sqrt{L})$ bytes of data transfer.

The servers' AES-NI encryption throughput limits the 
rate at which \name can process client requests at large table sizes.
Thus, the system's capacity to handle client write request scales
with the number of available CPU cores.
A large \name deployment could shard the database table across
$k$ machines to achieve a near-$k$-fold speedup.

We tested the system with anonymity set sizes of up to
2,895,216 clients, with a read-heavy latency-tolerant
microblogging workload.
To our knowledge, this is the largest anonymity set 
{\em ever constructed} in a system defending against 
traffic analysis attacks.
Prior DC-net-based systems scaled to 5,120 clients~\cite{wolinsky2012dissent}
and prior verifiable-shuffle-based systems 
scaled to 100,000 clients~\cite{bayer2012efficient}.
In contrast, \name scales to millions of clients for
certain applications.

\nicepara{Contributions}
This paper contributes:
\begin{compactitem}
  \item two new bandwidth-efficient and traffic-analysis-resistant 
        anonymous messaging protocols, 
        obtained by running private information retrieval 
        protocols ``in reverse'' (Sections~\ref{sec:arch} and~\ref{sec:dpf}),
  \item a fast method for excluding
        malformed client requests 
        (Section~\ref{sec:disrupt}), 
  \item a method to recover from transmission 
        collisions in DC-net-style anonymity systems, 
  \item experimental evaluation of these protocols with anonymity 
        set sizes of up to 2,895,216 users (Section~\ref{sec:eval}).
\end{compactitem}

\medskip
In Section~\ref{sec:goal}, we introduce our goals,
threat model, and security definitions.
Section~\ref{sec:arch} presents the high-level system
architecture. 
Section~\ref{sec:dpf} and Section~\ref{sec:disrupt}
detail our techniques for achieving bandwidth efficiency 
and disruption resistance in \name.
We evaluate the performance of the system in Section~\ref{sec:eval},
survey related work in Section~\ref{sec:rel}, and conclude
in Section~\ref{sec:concl}.

 \section{Goals and Problem Statement}
\label{sec:goal}

In this section, we summarize the high-level goals of the \name system and
present our threat model and security definitions.

\subsection{System Goals}

\Name implements an anonymous bulletin board using a
primitive we call a \textit{write-private database scheme}.
\Name enables clients to write into a shared
database, collectively maintained at a small set of servers,
without revealing to the servers the location or contents of the write.
Conceptually, the database table 
is just a long fixed-length bitstring divided into fixed-length rows. 

To write into the database, a client generates a {\em write request}. 
The write request encodes the message to be written
and the row index at which the client wants to write.
(A single client write request modifies a single 
database row at a time.)
Using cryptographic techniques, the client splits its write request
into a number of shares and the client sends 
one share to each of the servers.
By construction of the shares, no coalition of servers
smaller than a particular pre-specified threshold 
can learn the contents of a single client's write request.
While the cluster of servers must remain online for the duration 
of a protocol run, a client need only stay online for long enough to 
upload its write request to the servers.
As soon as the servers receive a write request, they
can apply it to to their local state. 

The \name cluster divides time into a series of epochs.
During each time epoch, servers collect many write requests from clients.
When the servers agree that the epoch has ended, they
combine their shares of the database to reveal the clients' plaintext messages.
A particular client's anonymity set consists of all of the honest clients
who submitted write requests to the servers during the time epoch.  
Thus, if 50,000 distinct honest clients submitted write requests
during a particular time epoch, each honest client is perfectly anonymous
amongst this set of 50,000 clients.

The epoch could be measured in time (e.g., 4 hours), in
a number of write requests (e.g., accumulate 10,000 write
requests before ending the epoch), or by
some more complicated condition
(e.g., wait for a write request signed from each of these
150 users identified by a pre-defined list of public keys).
The definition of what constitutes an epoch is crucial for security,
since a client's anonymity set is only as large as the number of
honest clients who submit write requests in the 
same epoch~\cite{serjantov2003trickle}.

When using \name as a platform for anonymous microblogging, the rows
would be long enough to fit a Tweet (140 bytes) and the number of rows would be
some multiple of the number of anticipated users.
To anonymously Tweet, a client would use the write-private database scheme to 
write its message into a random row of the database.
After many clients have written to the database, the servers can
reveal the clients' plaintext Tweets.
The write-privacy of the database scheme prevents eavesdroppers, malicious clients, 
and coalitions of malicious servers (smaller than a particular threshold) from
learning which client posted which message.

\subsection{Threat Model}
\label{sec:goal:threat}

Clients in our system are {\em completely untrusted}: they
may submit maliciously formed write requests to the system
and may collude with servers or with arbitrarily many other
clients to try to break the security properties of the system.

Servers in our system are trusted for availability. 
The failure---whether malicious or benign---of any one server 
renders the database state unrecoverable but {\em does not}
compromise the anonymity of the clients.
To protect against benign failures, server maintainers could
implement a single ``logical'' \name server with a cluster
of many physical servers running a
standard state-machine-replication 
protocol~\cite{liskov2012viewstamped,ongaro2014search}.

For each of the cryptographic instantiations of \name, there is a
threshold parameter $t$ that defines the number of malicious servers
that the system can tolerate while still maintaining its security
properties.  We make no assumptions about the behavior of malicious
servers---they can misbehave 
by publishing their secret keys, 
by colluding with coalitions of up
to $t$ malicious servers and arbitrarily many clients, 
or by mounting any other sort of attack against the system.

The threshold $t$ depends on the particular 
cryptographic primitives in use.
For our most secure scheme, {\em all but one} of the servers can
collude without compromising client privacy ($t = |{\sf Servers}| - 1$).
For our most efficient scheme, {\em no two} servers can collude ($t=1$).

\subsection{Security Goals}

The \name system implements a {\em write-private} and {\em disruption-resistant} 
database scheme. We describe the correctness and security properties for such 
a scheme here.

\begin{defn}[Correctness]
The scheme is {\em correct} if,
when all servers execute the protocol
faithfully, the plaintext state of the database revealed
at the end of a protocol run
is equal to the result of applying
each valid client write requests to an empty database
(i.e., a database of all zeros).
\end{defn}

Since we rely on all servers for availability,
correctness need only hold when all servers 
run the protocol correctly.

\medskip

To be useful as an anonymous bulletin board, the database
scheme must be {\em write-private} and {\em disruption resistant}.
We define these security properties here.

\nicepara{$(s,t)$-Write Privacy}
Intuitively, the system provides $(s,t)$-{\em write-privacy} if an 
adversary's advantage at guessing which honest
client wrote into a particular row of the database
is negligibly better than random guessing, even 
when the adversary controls all but two clients
and up to $t$ out of $s$ servers 
(where $t$ is a parameter of the scheme).
We define this property in terms of a {\em privacy game},
given in full in Appendix~\ref{app:game}.

\begin{defn}[$(s,t)$-Write Privacy]
We say that the protocol provides $(s,t)$-{\em write privacy} if the adversary's advantage
in the security game of Appendix~\ref{app:game}
is negligible in the (implicit) security parameter.
\end{defn}

\Name provides a very robust sort of privacy:
the adversary can select the messages that the honest clients will send and
can send maliciously formed messages that depend on the honest clients' messages.
Even then, the adversary still cannot guess which client uploaded which message.

\nicepara{Disruption resistance}
The system is {\em disruption resistant}
if an adversary who controls $n$ clients
can write into at most $n$ database rows during
a single time epoch.
A system that lacks disruption resistance might
be susceptible to denial-of-service attacks:
a malicious client could corrupt every row in the database
with a single write request.
Even worse, the write privacy of the system might prevent
the servers from learning which client was the disruptor.
Preventing such attacks is a major focus of prior
anonymous messaging schemes~\cite{chaum1988dining,goel2003herbivore,golle2004dining,waidner1989dining,wolinsky2012dissent}.
Under our threat model, we trust all servers for
availability of the system (though not for privacy).
Thus, our definition of disruption resistance concerns
itself only with clients attempting to disrupt the
system---we {\em do not} try to prevent servers
from corrupting the database state.

We formally define {\em disruption resistance} using
the following game, played between a challenger
and an adversary.
In this game, the challenger plays the role of
all of the servers and the adversary plays the 
role of all clients.
\begin{enumerate}
  \item The adversary sends $n$ write requests
    to the challenger
    (where $n$ is less than or equal to the number of 
    rows in the database).
  \item The challenger runs the protocol for
    a single time epoch, playing the role of the servers.
    The challenger then combines the servers' database
    shares to reveal the plaintext output.
\end{enumerate}

The adversary wins the game if the plaintext
output contains more than $n$ non-zero rows. 

\begin{defn}[Disruption Resistance]
We say that the protocol is {\em disruption resistant}
if the probability that the adversary wins the game
above is negligible in the (implicit) security parameter.
\end{defn}

\subsection{Intersection Attacks}

\Name makes it infeasible for an adversary to determine
which client posted which message {\em within} a particular time epoch.
If an adversary can observe traffic patterns {\em across} many
epochs, as the set of online clients changes, the adversary can
make statistical inferences about which client is sending 
which stream of 
messages~\cite{danezis2004statistical,kedogan2003limits,mathewson2005practical}.
These ``intersection'' or ``statistical disclosure'' attacks
affect many anonymity systems and defending against them is an important,
albeit orthogonal, problem~\cite{mathewson2005practical,wolinsky2013hang}.
Even so, intersection attacks typically become more difficult to
mount as the size of the anonymity set increases, so \name's support for
very large anonymity sets makes it less vulnerable to these attacks than 
are many prior systems.

 \section{System Architecture}
\label{sec:arch}

As described in the prior section,
a \name deployment consists of a small number
of servers, who maintain the database state, 
and a large number of clients.
To write into the database, a client
splits its write request using secret sharing
techniques and sends a single share to each of
the servers.
Each server updates its database state using
the client's share.
After collecting write requests from many
clients, the servers
combine their shares to reveal the plaintexts
represented by the write requests.
The security requirement is that 
no coalition of $t$ servers can learn which client
wrote into which row of the database.

\subsection{A First-Attempt Construction:\\Toy Protocol}
\label{sec:arch:straw}

As a starting point, we sketch a simple ``straw man''
construction that demonstrates the techniques
behind our scheme.
This first-attempt protocol shares some design features
with anonymous communication schemes based on 
client/server DC-nets~\cite{chaum1988dining,wolinsky2012dissent}.

In the simple scheme, we have two servers, $A$ and $B$,
and each server stores an $L$-bit bitstring, initialized to all zeros.
We assume for now that the servers {\em do not collude}---i.e.,
that one of the two servers is honest.
The bitstrings represent shares of the database state
and each ``row'' of the database is a single bit.

Consider a client who wants to write a ``1'' into 
row $\ell$ of the database.
To do so, the client generates a random $L$-bit bitstring $r$.
The client sends $r$ to server $A$ and $r \oplus e_\ell$ to server $B$,
where $e_\ell$ is an $L$-bit vector of zeros with a one at index~$\ell$
and $\oplus$ denotes bitwise XOR.
Upon receiving the write request from the client, each server XORs
the received string into its share of the database.

After processing $n$ write requests, the database state at
server $A$ will be:
\[ d_A = r_1 \oplus \dots \oplus r_n \]
and the database at server $B$ will be:
\begin{align*}
  d_B &= (e_{\ell_1} \oplus \dots \oplus e_{\ell_n}) \oplus (r_1 \oplus \dots \oplus r_n)\\
      &= (e_{\ell_1} \oplus \dots \oplus e_{\ell_n}) \oplus d_A
\end{align*}
At the end of the time epoch, the servers can reveal the plaintext
database by combining their local states $d_A$~and~$d_B$.

The construction generalizes to fields larger than $\F_2$.
For example, each ``row'' of the database could be a $k$-bit bitstring
instead of a single bit.
To prevent impersonation, network-tampering, and replay attacks, we 
use authenticated and encrypted channels with per-message nonces
bound to the time epoch identifier. 

\medskip

This protocol satisfies the write-privacy property
as long as the two servers do not collude (assuming that the clients
and servers deploy the replay attack defenses mentioned above).
Indeed, server $A$ can information theoretically simulate its view 
of a run of the protocol given only $e_{\ell_1} \oplus \dots \oplus e_{\ell_n}$ as input.
A similar argument shows that the protocol is write-private with
respect to server $B$ as well.

This first-attempt protocol has two major limitations.
The first limitation is that it is not bandwidth-efficient.
If millions of clients want to use the system in each time epoch, then 
the database must be at least millions of bits in length.
To flip a {\em single bit} in the database then, each client
must send {\em millions of bits} to each database,
in the form of a write request.

The second limitation is that it is not disruption resistant:
a malicious client can corrupt
the entire database with a single malformed request.
To do so, the malicious client picks random $L$-bit bitstrings
$r$ and $r'$, sends $r$ to server $A$, and sends $r'$ 
(instead of $r \oplus e_\ell$) to server $B$.
Thus, a single malicious client can efficiently
and anonymously deny service to all honest clients.

Improving bandwidth efficiency and adding
disruption resistance are the two core contributions
of this work, and we return to them in Sections~\ref{sec:dpf}
and~\ref{sec:disrupt}.

\subsection{Collisions} \label{sec:collisions}
Putting aside the issues of bandwidth efficiency and
disruption resistance for the moment, we now discuss
the issue of {\em colliding writes} to the shared database.
If clients write into random locations in the database,
there is some chance that one client's write request
will overwrite a previous client's message.
If client A writes message $m_A$ into location $\ell$,
client B might later write message $m_B$ into the
same location $\ell$.
In this case, row $\ell$ will contain $m_A \oplus m_B$, 
and the contents of row $\ell$ will be unrecoverable.

To address this issue,
we set the size of the database table to 
be large enough to accommodate the expected number 
of write requests for a given ``success rate.''
For example, the servers can choose a table size
that is large enough to accommodate $2^{10}$ write requests
such that $95\%$ of write requests
will not be involved in a collision (in expectation).
Under these parameters, $5\%$ of the write
requests will fail and those clients will have to
resubmit their write requests in a future time epoch.

We can determine the appropriate table size by solving
a simple ``balls and bins'' problem.
If we throw $m$ balls independently and uniformly at
random into $n$ bins, how many bins contain exactly one ball?
Here, the $m$ balls represent the write requests
and the $n$ bins represent the rows of the database.

Let $B_{ij}$ be the probability that ball $i$
falls into bin~$j$. For all $i$ and $j$, $\Pr[B_{ij}] = 1/n$.
Let $O_i^{(1)}$ be the event that {\em exactly} one ball falls
into bin~$i$.  Then
\[ \Pr\left[O_{i}^{(1)}\right] =  
                       \frac{m}{n}\left(1-\frac{1}{n}\right)^{m-1}  
\]
Expanding using the binomial theorem and ignoring low order terms
we obtain 
\[  \Pr\left[O_{i}^{(1)}\right] \approx         
                \frac{m}{n} -  \left(\frac{m}{n}\right)^2 + 
                \frac{1}{2}\left(\frac{m}{n}\right)^3
\]
where the approximation ignores terms of order $(m/n)^4$ and $o(1/n)$.
Then $n \cdot \Pr[O_{i}^{(1)}]$ is the expected number of bins
with exactly one ball which is the expected number of messages 
successfully received.  
Dividing this quantity by $m$ gives the expected success rate so that:
\[  \E[\text{SuccessRate}] = \frac{n}{m} \Pr[O_{i}^{(1)}] \approx 
      1 - \frac{m}{n} + \frac{1}{2} \left(\frac{m}{n}\right)^2  \]
So, if we want an expected success rate of $95\%$ then we need
$n \approx 19.5 m$.  For example, with $m = 2^{10}$ writers,
we would use a table of size $n \approx 20,000$.

\nicepara{Handling collisions}  We can shrink the table
size $n$ by coding the writes so that we can recover from collisions.
We show how to handle two-way collisions. 
That is, when at most two
clients write to the same location in the database.  Let us assume
that the messages being written to the database are elements in some
field $\F$ of odd characteristic (say $\F = \F_p$ where $p=2^{64}-59$).
We replace the XOR operation used in the basic scheme
by addition in $\F$.   

To recover from a two-way collision we will need to double the size of
each cell in the database, but the overall number of cells $n$ will
shrink by more than a factor of two.

When a client $A$ wants to write the message
$m_A \in \F$ to location~$\ell$ in the database the client will
actually write the pair $(m_A,m_A^2) \in \F^2$ into that location.  
Clearly if no
collision occurs at location $\ell$ then recovering $m_A$ at the end
of the epoch is trivial: simply drop the second coordinate (it is easy
to test that no collision occurred because the second coordinate is a
square of the first).  Now, suppose a collision occurs with some
client $B$ who also added her own message $(m_B,m_B^2) \in \F^2$ to
the same location~$\ell$ (and no other client writes to location
$\ell$).  Then at the end of the epoch the published values are
\[    S_1 = m_A + m_B  \pmod p \quad\text{and}\quad S_2 = m_A^2 + m_B^2  \pmod p\]
From these values it is quite easy to recover both $m_A$ and $m_B$
by observing that
\[  2 S_2 - S_1^2 = (m_A - m_B)^2  \pmod p\]
from which we obtain $m_A - m_B$ by taking a square root modulo $p$ (it does not
matter which of the two square roots we use---they both lead to the
same result).  Since $S_1 = m_A+m_B$ is also given it is now easy to
recover both $m_A$ and $m_B$.

Now that we can recover from two-way collisions we can shrink the
number of cells $n$ in the table.  Let $O_i^{(2)}$ be the event that 
exactly two balls fell into bin~$i$.  Then the expected number
of received messages is
\begin{equation} \label{eq:exp}
  n \Pr[O_{i}^{(1)}] + 2n \Pr[O_{i}^{(2)}]  
\end{equation}
where $\Pr[O_{i}^{(2)}] = {m \choose 2} \frac{1}{n^2} \left(1-\frac{1}{n}\right)^{m-2}$.
As before, dividing the expected number of received messages (\ref{eq:exp}) 
by $m$, expanding using the binomial theorem, and ignoring low order terms
gives the expected success rate as:
\[    \E[\text{SuccessRate}] 
                \approx 
      1 - \frac{1}{2} \left(\frac{m}{n}\right)^2 + 
           \frac{1}{3} \left(\frac{m}{n}\right)^3  \]
So, if we want an expected success rate of $95\%$ 
we need a table with $n \approx 2.7 m$ cells.  
This is a far smaller table than before, when we could not handle 
collisions.  In that case we needed $n \approx 19.5 m$ which results
in much bigger tables, despite each cell being half as big.
Shrinking the table reduces the storage
and computational burden on the servers.

\medskip
This two-way collision handling technique generalizes to handle
$k$-way collisions for $k>2$.
To handle $k$-way collisions, we increase the size of each cell by a
factor of $k$ and have each client $i$ write $(m_i, m_i^2, \ldots,
m_i^k) \in \F^k$ to its chosen cell.  A $k$-collision gives $k$
equations in $k$ variables that can be efficiently solved to recover
all $k$ messages, as long as the characteristic of $\F$ is greater
than $k$~\cite{bos1989detection,chien1966application}.
Using $k>2$ further reduces the table size as the
desired success rate approaches one.

The collision handling method described in this section will also
improve performance of our full system, which we describe in the next section.

\nicepara{Adversarial collisions} The analysis above assumes that
clients behave honestly.  Adversarial clients, however,
need not write into random rows of the database---i.e.,
all $m$ balls might not be thrown independently and uniformly at
random.  A coalition of clients might, for example, try to increase the
probability of collisions by writing into the database using some
malicious strategy.

By symmetry of writes we can assume that all $\hat{m}$ adversarial
clients write to the database before the honest clients do.  
Now a message from an honest client is properly received at the end of an
epoch if it avoids all the cells filled by the malicious
clients.  We can therefore carry out the honest client analysis above
assuming the database contain $n-\hat{m}$ cells instead of $n$ cells.
In other words, given a bound $\hat{m}$ on the number of malicious
clients we can calculate the required table size $n$.  In practice, if
too many collisions are detected at the end of an epoch the servers can
adaptively double the size of the table so that the next epoch has
fewer collisions.

\subsection{Forward Security}
\label{sec:arch:forward}

Even the first-attempt scheme
sketched in Section~\ref{sec:arch:straw} provides
{\em forward security} in the event that all
of the servers' secret keys are compromised~\cite{canetti2003forward}.
To be precise: an adversary could compromise the
state and secret keys of {\em all servers} after the servers
have processed $n$ write requests from honest clients, 
but {\em before} the time epoch has ended.
Even in this case, 
the adversary will be unable to determine which of the
$n$ clients submitted which of the $n$ plaintext
messages with a non-negligible advantage over random guessing.
(We assume here that clients and servers
communicate using encrypted channels which themselves
have forward secrecy~\cite{rfc7296}.)

This forward security property means that clients need not
trust that $S-t$ servers stay honest forever---just that they
are honest at the moment when the client submits its upload request.
Being able to weaken the trust assumption about the servers
in this way might be valuable in hostile environments, in which
an adversary could compromise a server at any time without warning.

Mix-nets do not have this property, since servers must accumulate
a set of onion-encrypted messages before shuffling and
decrypting them~\cite{chaum1981untraceable}. 
If an adversary always controls the first mix server
and if it can compromise the rest of the mix servers after
accumulating a set of ciphertexts,
the adversary can de-anonymize all of the system's users.
DC-net-based systems that use ``blame'' protocols to 
retroactively discover disruptors have a similar 
weakness~\cite{corrigangibbs2010dissent,wolinsky2012dissent}.

The full \name protocol
maintains this forward security property.

 \section{Improving Bandwidth Efficiency with Distributed Point Functions}
\label{sec:dpf}

This section describes how application of private
information retrieval techniques can improve the bandwidth
efficiency of the first-attempt protocol.

\nicepara{Notation}
The symbol $\F$ denotes an arbitrary finite field,
$\Z_L$ is the ring of integers modulo $L$.
We use $e_{\ell} \in \F^L$ to represent a vector 
that is zero everywhere except at index $\ell \in \Z_L$,
where it has value ``$1$.''
Thus, for $m \in \F$, the vector $m \cdot e_\ell \in \F^L$ is the 
vector whose value is zero everywhere except at index $\ell$, where it
has value $m$.
For a finite set $S$, the notation $x \rgets S$ indicates
that the value of $x$ is sampled independently and uniformly at random
from $S$.
The element ${\bf v}[i]$ is the value of
a vector ${\bf v}$ at index $i$.
We index vectors starting at zero.

\subsection{Definitions}
The bandwidth inefficiency of the protocol sketched above comes from the
fact that the client must send an $L$-bit vector
to each server to flip a single bit in the logical database.
To reduce this $O(L)$ bandwidth overhead, we apply techniques inspired
by private information retrieval 
protocols~\cite{chor1997computationally,chor1998private,gilboa2014distributed}.

The problem of private information retrieval (PIR) is essentially the converse
of the problem we are interested in here.
In PIR, the client must {\em read} 
a bit from a replicated database without revealing to 
the servers the index being read.
In our setting, the client must {\em write}
a bit into a replicated database without revealing to the servers the
index being written.
Ostrovsky and Shoup first made this
connection in the context of a ``private information storage'' protocol~\cite{ostrovsky1997private}.

PIR schemes allow the client to split its query to the servers 
into shares such that  
(1) a subset of the shares does not leak information about the index of interest, and
(2) the length of the query shares is much less than the length of the database.
The core building block of many PIR schemes, which we adopt
for our purposes, is a {\em distributed point function}.
Although Gilboa and Ishai~\cite{gilboa2014distributed}
defined distributed point functions 
as a primitive only recently, many prior PIR schemes make implicit
use the primitive~\cite{chor1997computationally,chor1998private}.
Our definition of a distributed point function
follows that of Gilboa and Ishai, except
that we generalize the definition to allow 
for more than two servers.

First, we define a (non-distributed) point function.

\begin{defn}[Point Function]
Fix a positive integer $L$ and a finite field $\F$.
For all $\ell \in \Z_L$ and $m \in \F$, the {\em point function} $P_{\ell,m}: \Z_L \to \F$
is the function such that $P_{\ell,m}(\ell) = m$ and $P_{\ell,m}(\ell') = 0$ for
all $\ell \neq \ell'$.
\end{defn}

That is, the point function $P_{\ell,m}$ has the value $0$ when evaluated at any 
input not equal to $\ell$ and it has the value $m$ when evaluated at $\ell$.
For example, if $L = 5$ and $\F = \F_2$, the point function 
$P_{3,1}$ takes on the values $(0, 0, 0, 1, 0)$ when evaluated
on the values $(0, 1, 2, 3, 4)$ (note that we index vectors from zero).

An $(s,t)$-distributed point function provides a way to 
distribute a point function $P_{\ell,m}$ amongst $s$ servers such that
no coalition of at most $t$ servers learns anything about $\ell$ or $m$
given their $t$ shares of the function.

\begin{defn}[Distributed Point Function (DPF)]
Fix a positive integer $L$ and a finite field $\F$.
An {\em $(s,t)$-distributed point function} consists of a pair
of possibly randomized algorithms that implement the following functionalities:
\begin{compactitem}

\item
  $\mathsf{Gen}(\ell, m) \rightarrow (k_0, \dots, k_{s-1})$.
Given an integer $\ell \in \Z_L$ and value $m \in \F$, 
output a list of $s$ keys.

\item
$\mathsf{Eval}(k, \ell') \rightarrow m'$.
Given a key $k$ generated using ${\sf Gen}$, and 
an index $\ell' \in \Z_L$, return a value $m' \in \F$.
\end{compactitem}
\end{defn}

We define correctness and privacy for 
a distributed point function as follows:
\begin{compactitem}

\item
{\bf Correctness.}
For a collection of $s$ keys generated using ${\sf Gen}(\ell,m)$,
the sum of the outputs of these keys (generated using ${\sf Eval}$)
must equal the point function $P_{\ell,m}$.
More formally, for all $\ell, \ell' \in \Z_L$ and $m \in \F$:
\begin{multline*} 
\Pr[(k_0, \dots, k_{s-1}) \gets \mathsf{Gen}(\ell, m) :\\ 
    \Sigma_{i=0}^{s-1} \mathsf{Eval}(k_i, \ell') = P_{\ell,m}(\ell')] = 1 
\end{multline*}
where the probability is taken over the randomness of the ${\sf Gen}$ algorithm.

\item
{\bf Privacy.}
Let $S$ be any subset of $\{0, \dots, s-1\}$ 
such that $|S| \leq t$.
Then for any $\ell \in \Z_L$ and $m \in \F$,
let $D_{S,\ell,m}$ denote the distribution of
keys $\{ (k_i)\ |\ i \in S\}$ induced by 
$(k_0, \dots, k_{s-1}) \gets {\sf Gen}(\ell, m)$.
We say that an $(s,t)$-DPF maintains privacy if
there exists a p.p.t. algorithm ${\sf Sim}$ such that
the following distributions are computationally
indistinguishable:
\[ D_{S,\ell,m} \approx_c {\sf Sim}(S) \]
That is, any subset of at most
$t$ keys leaks no information about $\ell$ or $m$.
(We can also strengthen this definition to require statistical
or perfect indistinguishability.)
\end{compactitem}

\medskip

\nicepara{Toy Construction}
To make this definition concrete,
we first construct a trivial information-theoretically secure 
\mbox{$(s, s-1)$}-distributed point function with length-$L$ keys.
As above, we fix a length $L$ and a finite field $\F$.

\begin{compactitem}
\item
$\mathsf{Gen}(\ell, m) \rightarrow (k_0, \dots, k_{s-1})$.
Generate random vectors $k_0, \dots, k_{s-2} \in \F^L$.
Set $k_{s-1} = m \cdot e_{\ell} - \Sigma_{i=0}^{s-2} k_i$.

\item
$\mathsf{Eval}(k, \ell') \rightarrow m'$.
Interpret $k$ as a vector in $\F^L$.
Return the value of the vector $k$ at index $\ell'$.

\end{compactitem}
The correctness property of this
construction follows immediately.
Privacy is maintained because the distribution of
any collection of $s-1$ keys is independent of $\ell$ and~$m$.

This toy construction uses length-$L$ keys to distribute
a point function with domain $\Z_L$.
Later in this section we describe 
DPF constructions which use much shorter keys.

\subsection{Applying Distributed Point Functions for Bandwidth Efficiency}

We can now use DPFs to improve the efficiency of the
write-private database scheme introduced in Section~\ref{sec:arch:straw}.
We show that the existence of an $(s,t)$-DPF with keys of length 
$|k|$ (along with standard cryptographic assumptions) implies
the existence of write-private database scheme using $s$ servers
that maintains anonymity in the presence of $t$ malicious servers,
such that write requests have length $s|k|$.
Any DPF construction with short keys thus immediately implies 
a bandwidth-efficient write-private database scheme.

The construction is a generalization of the one presented
in Section~\ref{sec:arch:straw}.
We now assume that there are $s$ servers such that no more than $t$
of them collude.
Each of the $s$ servers maintains a vector in $\F^L$ as their
database state, for some fixed 
finite field $\F$ and integer $L$.
Each ``row'' in the database is now an element of $\F$ and the database
has $L$ rows.

When the client wants to write a message $m \in \F$ into location
$\ell \in \Z_L$ in the database, the client 
uses an $(s,t)$-distributed point function to generate a set of
$s$ DPF keys:
\[ (k_0, \dots, k_{s-1}) \gets {\sf Gen}(\ell, m) \]
The client then sends one of the keys to each of the servers.
Each server $i$ can then expand
the key into a vector $v \in \F^L$
by computing $v(\ell') = {\sf Eval}(k_i, \ell')$ for $\ell' = 0, \dots, L-1$.
The server then adds this vector $v$ into its database state,
using addition in $\F^L$.
At the end of the time epoch, all servers combine their database
states to reveal the set of client-submitted messages.

\nicepara{Correctness}
The correctness of this construction follows directly
from the correctness of the DPF.
For each of the $n$ write requests submitted by the clients, 
denote the $j$-th key in the $i$-th request as $k_{i,j}$,
denote the write location as $\ell_i$, and the message
being written as $m_i$.
When the servers combine their databases at the end of the
epoch, the contents of the final database at row $\ell$ will be:
\begin{align*}
d_\ell = \sum_{i=0}^{n-1} \sum_{j=0}^{s-1} {\sf Eval}(k_{i,j}, \ell) 
    = \sum_{i=0}^{n-1} P_{\ell_i, m_i}(\ell) \quad \in \F
\end{align*}
In words: as desired, 
the combined database contains the sum of $n$ point
functions---one for each of the write requests.

\nicepara{Anonymity}
The anonymity of this construction 
follows directly from the privacy property of the DPF.
Given the plaintext database state $d$ (as defined above),
any coalition of $t$ servers can simulate its view of the protocol.
By definition of DPF privacy, there exists
a simulator ${\sf Sim}$, which simulates the
distribution of any subset of $t$ DPF keys generated
using $\Gen$.
The coalition of servers can use this simulator to simulate each
of the $n$ write requests it sees during a run of the protocol.
Thus, the servers can simulate their view of a
protocol run and cannot win the anonymity game
with non-negligible advantage. 

\nicepara{Efficiency}
A client in this scheme sends $|k|$ bits to each
server (where $k$ is a DPF key), so the 
bandwidth efficiency of 
the scheme depends on the efficiency of the DPF.
As we will show later in this section, $|k|$ can be
much smaller than the length of the database.

\subsection{A Two-Server Scheme Tolerating One Malicious Server}
\label{sec:dpf:twoserver}

\begin{figure*}
\centering
\includegraphics[width=\textwidth]{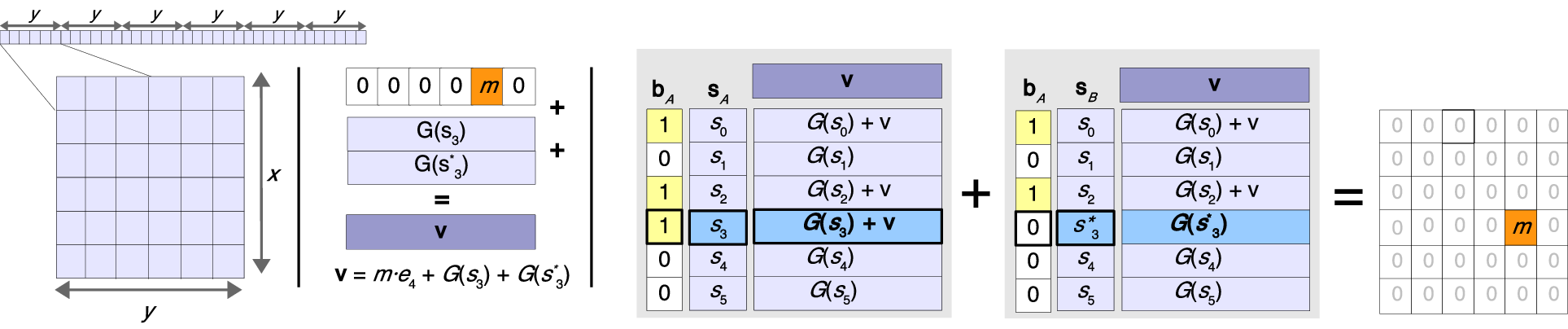}
\caption{Left: We represent the 
output of $\Eval$ as an $x \times y$ matrix of field
elements.
Left-center: Construction of the ${\bf v}$ vector used
in the DPF keys.
Right: using the ${\bf v}$, ${\bf s}$, and ${\bf b}$ vectors,
$\Eval$ expands each of the two keys into an 
$x\times y$ matrix of field elements.
These two matrices sum to zero everywhere except 
at $(\ell_x,\ell_y) = (3,4)$, where they sum to $m$.
}
\label{fig:dpf}
\end{figure*}

Having established that DPFs with short keys lead to 
bandwidth-efficient write-private database schemes,
we now present one such DPF construction. 
This construction is a simplification of
computational PIR scheme of Chor and 
Gilboa~\cite{chor1997computationally}.

This is a $(2,1)$-DPF with keys of length $O(\sqrt{L})$
operating on a domain of size $L$.
This DPF yields a two-server write-private database scheme 
tolerating one malicious server such that writing into a database
of size $L$ requires sending $O(\sqrt{L})$ 
bits to each server.
Gilboa and Ishai~\cite{gilboa2014distributed} construct
a $(2,1)$-DPF with even shorter keys ($|k| = \mathsf{polylog}(L)$),
but the construction presented here is efficient enough for the database sizes
we use in practice.
Although the DPF construction works over any field, we
describe it here using the binary field $\F = \F_{2^k}$ 
(the field of $k$-bit bitstrings) to simplify the exposition.

When $\Eval(k, \ell')$ is run on every
integer $\ell' \in \{0, \dots, L-1\}$, 
its output is a vector of $L$ field elements.
The DPF key construction conceptually works by representing this
a vector of $L$ field elements
as an $x \times y$ matrix, such that $xy \geq L$.
The trick that makes the construction work is that the size of the keys
needs only to grow with the size of the {\em sides} of this matrix
rather than its {\em area}.
The DPF keys that $\Gen(\ell, m)$ outputs give an efficient
way to construct two matrices $M_A$ and $M_B$ that differ only 
at one cell $\ell = (\ell_x, \ell_y) \in \Z_x \times \Z_y$
(Figure~\ref{fig:dpf}).

Fix a binary finite field $\F = \F_{2^k}$, 
a DPF domain size $L$, and integers $x$ and $y$ such that $xy \geq L$.
(Later in this section, we describe how to 
choose $x$ and $y$ to minimize the key size.)
The construction requires a pseudo-random generator (PRG)
$G$ that stretches seeds from some space $\Seed$ 
into length-$y$ vectors of elements of $\F$~\cite{haastad1999pseudorandom}.
So the signature of the PRG is $G: \Seed \rightarrow \F^y$.
In practice, an implementation might use AES-128 
in counter mode as the pseudo-random generator~\cite{nist2001aes}.

The algorithms comprising the DPF are:
\begin{compactitem}
\item
$\mathsf{Gen}(\ell, m) \rightarrow (k_A, k_B)$.
Compute integers $\ell_x \in \Z_x$ and $\ell_y \in \Z_y$ 
such that $\ell = \ell_x y + \ell_y$.
Sample a random bit-vector ${\bf b}_A \rgets \{0,1\}^x$,
a random vector of PRG seeds ${\bf s}_A \rgets \Seed^x$,
and a single random PRG seed $s_{\ell_x}^* \rgets \Seed$.

Given ${\bf b}_A$ and ${\bf s}_A$, we define 
${\bf b}_B$ and ${\bf s}_B$ as: 
\begin{align*}
  {\bf b}_A = (b_0, \dots, b_{\ell_x}, \dots, b_{x-1}) \\
      {\bf b}_B = (b_0, \dots, \bar{b}_{\ell_x}, \dots, b_{x-1}) \\
  {\bf s}_A = (s_0, \dots, s_{\ell_x}, \dots, s_{x-1}) \\
      {\bf s}_B = (s_0,\dots, s^*_{\ell_x}, \dots, s_{x-1}) 
\end{align*}
That is, the vectors ${\bf b}_A$ and ${\bf b}_B$
(similarly ${\bf s}_A$ and ${\bf s}_B$) 
differ only at index $\ell_x$.

Let $m \cdot e_{\ell_y}$ be the vector in $\F^y$ of all zeros except
that it has value $m$ at index $\ell_y$.
Define ${\bf v} \gets m \cdot e_{\ell_y} + G(s_{\ell_x}) + G(s_{\ell_x}^*)$.

The output DPF keys are:
\[ k_A = ({\bf b}_A, {\bf s}_A, {\bf v}) \qquad 
    k_B = ({\bf b}_B, {\bf s}_B, {\bf v}) \]

\item
$\mathsf{Eval}(k, \ell') \rightarrow m'$.
Interpret $k$ as a tuple $({\bf b}, {\bf s}, {\bf v})$.
To evaluate the PRF at index $\ell'$, 
first write $\ell'$ as an $(\ell'_x,\ell'_y)$ tuple such that
$\ell_x' \in \Z_x$, $\ell_y' \in \Z_y$, and $\ell' = \ell'_x y + \ell'_y$.
Use the PRG $G$ to stretch the $\ell'_x$-th seed
of ${\bf s}$ into a length-$y$ vector: 
${\bf g} \gets G({\bf s}[\ell'_x])$.
Return $m' \gets ({\bf g}[\ell'_y] + {\bf b}[\ell'_x]{\bf v}[\ell'_y])$.

\end{compactitem}

Figure~\ref{fig:dpf} graphically depicts
how ${\sf Eval}$ stretches the 
keys into a table of $x \times y$ field elements.

\nicepara{Correctness}
We prove correctness of the scheme in Appendix~\ref{app:correct}.

\nicepara{Privacy}
The privacy property requires that there exists
an efficient simulator that, on input ``$A$'' or ``$B$,''
outputs samples from a distribution that is computationally
indistinguishable from the distribution of DPF keys
$k_A$ or $k_B$.

The simulator ${\sf Sim}$ simulates each component of the
DPF key as follows:
It samples 
${\bf b} \rgets \{0,1\}^x$,
${\bf s} \rgets \Seed^x$, and
${\bf v} \rgets \F^y$. 
The simulator returns $({\bf b}, {\bf s}, {\bf v})$.

We must now argue that the simulator's output distribution
is computationally indistinguishable from that induced by the
distribution of a single output of \Gen.
Since the ${\bf b}$ and ${\bf s}$ vectors outputted by 
\Gen are random, the simulation is perfect.
The ${\bf v}$ vector outputted by \Gen is computationally indistinguishable
from random, since it is padded with the output of the PRG seeded with 
a seed unknown to the holder of the key.
An efficient algorithm
to distinguish the simulated ${\bf v}$ vector from 
random can then also distinguish the PRG output from random.

\nicepara{Key Size}
A key for this DPF scheme consists of:
a vector in $\{0,1\}^x$,
a vector in $\Seed^x$, and
a vector in $\F^y$.
Let $\alpha$ be the number of bits required to represent
an element of $\Seed$ and let $\beta$ be the number of
bits required to represent an element of $\F$.
The total length of a key is then:
\[ |k| = (1 + \alpha) x + \beta y \]
For fixed spaces $\Seed$ and $\F$, we can find the optimal
choices of $x$ and $y$ to minimize the key length.
To do so, we solve:
\[ \min_{x,y} ((1+\alpha) x + \beta y) \quad 
    \textrm{ subject to } \quad xy \geq L \]
and conclude that the optimal values of $x$ and $y$ are:
\[ x = c\sqrt{L} \quad \textrm{and} \quad
    y = \frac{1}{c}\sqrt{L}\quad \textrm{where} \quad
    c = \sqrt{\frac{\beta}{1+\alpha}}.\]
The key size is then $O(\sqrt{L})$.

When using a database table of 
one million rows in length ($L = 2^{20}$),
a row length of 1 KB per row ($\F = \F_{2^{8192}}$), 
and a PRG seed size of 128 bits (using AES-128, for example)
the keys will be roughly 263 KB in length.
For these parameters, 
the keys for the na\"ive construction
(Section~\ref{sec:arch:straw})
would be 1 GB in length.
Application of efficient DPFs thus yields
a 4,000$\times$ bandwidth savings in this case.

\nicepara{Computational Efficiency}
A second benefit of this scheme is that
both the \Gen and ${\sf Eval}$ routines are 
computationally efficient, since they just require performing
finite field additions (i.e., XOR for binary fields)
and PRG operations (i.e., computations of the AES function).
The construction requires no public-key primitives. 

\subsection{An $s$-Server Scheme Tolerating $s-1$ Malicious Servers}
\label{sec:dpf:manyserver}

The $(2,1)$-DPF scheme described above 
achieved a key size of $O(\sqrt{L})$ bits using only 
symmetric-key primitives.
The limitation of that construction is that it only
maintains privacy when a single key is compromised.
In the context of a write-private database scheme, this
means that the construction 
can only maintain anonymity in the presence
of a {\em single} malicious server.
It would be much better to have a write-private database 
scheme with $s$ servers that maintains anonymity in the
presence of $s-1$ malicious servers.
To achieve this stronger security notion, we need a
bandwidth-efficient
$(s,s-1)$-distributed point function.

In this section, we construct an $(s,s-1)$-DPF
where each key has size $O(\sqrt{L})$.
We do so at the cost of requiring more expensive 
public-key cryptographic operations, 
instead of the symmetric-key operations used
in the prior DPF.
While the $(2,1)$-DPF construction above
directly follows the work of Chor and Gilboa~\cite{chor1997computationally},
this $(s,s-1)$-DPF construction is novel, as far as we know.
In recent work, Boyle et al.~present a $(s,s-1)$-DPF construction using only symmetric-key 
operations, but this construction exhibits a key size {\em exponential} in 
the number of servers $s$~\cite{boyle2015function}.

\medskip

This construction uses
a \textit{seed-homomorphic pseudorandom 
generator}~\cite{banerjee2014new,boneh2013key,naor1999distributed},
to split the key for the pseudo-random generator $G$ across
a collection of $s$ DPF keys.

\begin{defn}[Seed-Homomorphic PRG]
A {\em seed-homomorphic} PRG is a pseudo-random generator
$G$ mapping seeds in a group
$(\Seed, \oplus)$ to outputs in a group $(\G, \otimes)$ 
with the additional property that for any $s_0, s_1 \in \Seed$:
\[ G(s_0 \oplus s_1) = G(s_0) \otimes G(s_1) \]
\end{defn}

It is possible to construct a simple seed-homomorphic PRG from the decision
Diffie-Hellman (DDH) assumption~\cite{boneh2013key,naor1999distributed}.
The public parameters for the scheme are 
list of $y$ generators chosen at random from an order-$q$ 
group $\G$, in which the DDH problem is hard~\cite{boneh1998decision}.
For example, if $\G$ is an elliptic curve group~\cite{miller1986use},
then the public parameters will be $y$ points $(P_0, \dots, P_{y-1}) \in \G^y$.
The seed space is $\Z_q$ and the generator
outputs vectors in $\G^y$.
On input $s \in \Z_q$, the generator outputs
$(sP_0, \dots, sP_{y-1})$.
The generator is seed-homomorphic because,
for any $s_0,s_1 \in \Z_q$, and for
all $i \in \{1, \dots, y\}$: 
$s_0P_i + s_1 P_i = (s_0+s_1)P_i$. 

\medskip

As in the prior DPF construction, we fix
a DPF domain size $L$, and
integers $x$ and $y$ such that $xy \geq L$.
The construction requires a seed-homomorphic PRG
$G: \Seed \mapsto \G^y$, for some group $\G$
of prime order~$q$.

For consistency with the prior DPF construction,
we will write the group operation in $\G$ 
using additive notation.
Thus, the group operation applied component-wise
to vectors ${\bf u}, {\bf v} \in \G^y$ results in
the vector $({\bf u} + {\bf v}) \in \G^y$.
Since $\G$ has order $q$, $qA = 0$ for all $A \in \G$.

The algorithms comprising the $(s,s-1)$-DPF are:
\begin{compactitem}
\item
$\mathsf{Gen}(\ell, m) \rightarrow (k_0, \dots, k_{s-1})$.
Compute integers $\ell_x \in \Z_x$ and $\ell_y \in \Z_y$ 
such that $\ell = \ell_x y + \ell_y$.
Sample random integer-valued vectors ${\bf b}_0, \dots, {\bf b}_{s-2} \rgets (\Z_q)^x$,
random vectors of PRG seeds ${\bf s}_0, \dots, {\bf s}_{s-2} \rgets \Seed^x$,
and a single random PRG seed $s^* \rgets \Seed$.

Select ${\bf b}_{s-1} \in (\Z_q)^x$ such that 
$\Sigma_{k=0}^{s-1} {\bf b}_k = e_{\ell_x} \pmod q$
and select
${\bf s}_{s-1} \in \Seed^x$ such that 
$\Sigma_{k=0}^{s-1} {\bf s}_k = s^* \cdot e_{\ell_x} \in \G^x$.
Define ${\bf v} \gets m \cdot e_{\ell_y} - G(s^*)$.

The DPF key for server $i \in \{0, \dots, s-1\}$ is
$k_i = ({\bf b}_i, {\bf s}_i, {\bf v})$.

\item
$\mathsf{Eval}(k, \ell') \rightarrow m'$.
Interpret $k$ as a tuple $({\bf b}, {\bf s}, {\bf v})$.
To evaluate the PRF at index $\ell'$, 
first write $\ell'$ as an $(\ell'_x,\ell'_y)$ tuple such that
$\ell'_x \in \Z_x$, $\ell'_y \in \Z_y$, and $\ell' = \ell'_x y + \ell'_y$.
Use the PRG $G$ to stretch the $\ell'_x$-th seed
of ${\bf s}$ into a length-$y$ vector: 
${\bf g} \gets G({\bf s}[\ell'_x])$.
Return $m' \gets ({\bf g}[\ell'_y] + {\bf b}[\ell'_x]{\bf v}[\ell'_y])$.

\end{compactitem}

We omit correctness and privacy proofs, since they follow exactly the same
structure as those used to prove security of our prior DPF construction.
The only difference is that correctness here
relies on the fact that $G$ is a seed-homomorphic PRG, rather than a
conventional PRG.
As in the DPF construction
of Section~\ref{sec:dpf:twoserver},
the keys here are of length
$O(\sqrt{L})$.

\nicepara{Computational Efficiency}
The main computational cost of this DPF construction
comes from the use of the seed-homomorphic PRG $G$.
\textit{Unlike} a conventional PRG, which can be implemented
using AES or another fast block cipher in counter mode,
known constructions of seed-homomorphic PRGs require
algebraic groups~\cite{naor1999distributed} 
or lattice-based cryptography~\cite{banerjee2014new,boneh2013key}.

When instantiating the $(s,s-1)$-DPF with 
the DDH-based PRG construction in elliptic curve groups,
each call to the DPF ${\sf Eval}$ routine 
requires an expensive 
elliptic curve scalar multiplication.
Since elliptic curve operations are, per byte,
orders of magnitude slower than AES operations,
this $(s,s-1)$-DPF will be orders
of magnitude slower than the $(2,1)$-DPF.
Security against an arbitrary number of malicious servers
comes at the cost of computational efficiency, at least
for these DPF constructions.

\medskip

With DPFs, we can now construct
a bandwidth-efficient write-private database scheme
that tolerates one malicious server (first construction)
or $s-1$ out of $s$ malicious servers (second construction).

 \section{Preventing Disruptors}
\label{sec:disrupt}

The first-attempt construction of
our write-private database scheme (Section~\ref{sec:arch:straw})
had two limitations: 
(1) client write requests were very large 
and (2) malicious clients could corrupt the database state
by sending malformed write requests.
We addressed the first of these two challenges
in Section~\ref{sec:dpf}.
In this section, we address the second challenge.

A client write request in our protocol just consists
of a collection of $s$ DPF keys.
The client sends one key to each of the $s$ servers.
The servers must collectively decide whether the collection
of $s$ keys is a valid output of the DPF ${\sf Gen}$ routine,
without revealing any information about the keys themselves.

One way to view the servers' task here is as a secure
multi-party computation~\cite{goldreich1987play,yao1982protocols}.
Each server $i$'s private input is its DPF key $k_i$.
The output of the protocol is a single bit, which determines if the $s$ keys $(k_0, \dots, k_{s-1})$ are a
well-formed collection of DPF keys.

Since we already rely on servers for availability
(Section~\ref{sec:goal:threat}), we {\em need not} protect
against servers maliciously trying to manipulate the 
output of the multi-party protocol.
Such manipulation could only result in corrupting the database 
(if a malicious server accepts a write request that it should
have rejected)
or denying service to an honest client
(if a malicious server rejects a write request that it should
have accepted).
Since both attacks are tantamount to denial of service,
we need not consider them.

We {\em do} care, in contrast, about protecting 
client privacy against malicious servers.
A malicious server participating in the protocol
should not gain any additional information about
the private inputs of other parties, no matter how
it deviates from the protocol specification.

We construct two protocols for checking the validity
of client write requests.
The first protocol is computationally inexpensive, but 
requires introducing a third non-colluding party to 
the two-server scheme. 
The second protocol requires relatively expensive zero-knowledge proofs~\cite{feige1988zero,goldreich1991proofs,goldwasser1989knowledge,rackoff1992non},
but it maintains security when all but one
of $s$ servers is malicious.
Both of these protocols must satisfy the 
standard notions of soundness, completeness, and
zero-knowledge~\cite{camenisch1998group}.

Since the publication of this paper, Boyle et al.~have designed a very
efficient protocol for checking the well-formedness of DPF
keys~\cite{boyle2016function}.
Their checking protocol provides security against semi-honest (i.e., honest but curious) servers
and requires only a \textit{constant} amount of server-to-server communication.
If it is possible to extend their checking protocol to provide security against fully malicious
servers, their new scheme could serve as a more efficient alternative to the protocols described herein.

\subsection{Three-Server Protocol}
\label{sec:disrupt:smc}

Our first protocol for detecting malformed write requests
works with the $(2,1)$-DPF scheme presented in 
Section~\ref{sec:dpf:twoserver}.
The protocol uses only hashing and finite field additions, 
so it is computationally inexpensive.
The downside is that it requires introducing a third
{\em audit} server, which must not collude with either
of the other two servers.
This simple protocol draws inspiration from classical secure
multi-party computation protocols~\cite{fagin1996comparing,
goldreich1987play, yao1982protocols}.

As a warm-up to the full checking protocol, 
we develop a three-server protocol called
${\sf AlmostEqual}$
that we use as a subroutine 
to implement the full write-request validation protocol.
The ${\sf AlmostEqual}$ protocol takes place between 
the client and three servers: 
database server $A$, database server $B$, and an audit server.

Let $\lambda$ be a security parameter (e.g., $\lambda = 256$).  
The protocol uses a hash function 
$H: \{0,1\}^* \to \{0,1\}^{\lambda}$, which we model as a random
oracle~\cite{bellare1993random}.

At the start of the protocol, database server $A$ holds 
a vector ${\vA \in \F^n}$, and database server $B$ holds a vector
${\vB \in \F^n}$.
The audit server takes no private input.
The client holds both $\vA$ and $\vB$.

The three servers execute a protocol that allows them to confirm
that the vectors $\vA$ and $\vB$ are equal everywhere except at exactly one index, and
such that any one malicious server learns nothing about 
the index at which these vectors differ,
as long as the vectors indeed differ at a single index.

More formally, the servers want to execute this 
check in such a way that the following properties hold:
\begin{compactitem}
\item[--] \textbf{Completeness.} If all parties are honest, and
    $\vA$ and $\vB$ are well formed, then the database servers 
    almost always accept the vectors as well formed.
\item[--] \textbf{Soundness.} If $\vA$ and $\vB$ do not differ
    at a single index, and all three servers are honest, then the
    servers will reject the vectors almost always.
\item[--] \textbf{Zero knowledge.} If $\vA$ and $\vB$ are well formed
    and the client is honest, then any one actively malicious server
    can simulate its view of the protocol execution. Furthermore,
    the simulator does not take as input the index at which $\vA$ and $\vB$
    differ nor the value of the vectors at that index.
\end{compactitem}

We denote an instance of the three-server
protocol as ${\sf AlmostEqual}({\bf v}_A, {\bf v}_B)$,
where the arguments denote the private values that 
the two database servers take as input.
The protocol proceeds as follows:
\begin{enumerate}
  \item The client sends a PRG seed $\sigma \in \lbits$ to both database servers.

  \item Servers $A$ and $B$ use a PRG seeded with the seed $\sigma$
        to sample $n$ pseudorandom values $(r_0, \dots, r_{n-1}) \in \{0,1\}^{\lambda n}$. 
        The servers also use the seed $\sigma$ to agree upon
        a pseudorandom ``shift'' value $f \in \Z_n$.
        
  \item Server $A$ computes the values $m_i \gets H({\bf v}_A[i], r_i)$
        for every index $i \in \{0, \dots, n-1\}$ and 
        sends to the auditor
        \[{\bf m}_A = (m_f, m_{f+1}, \dots, m_{n-1}, m_0, \dots, m_{f-1}).\] 
        The vector $\mA$ is a blinded version of server $A$'s input vector $\vA$.
        Using a secret random value $\rho \in \{0,1\}^\lambda$ shared with server $B$
        (constructed using a coin-flipping protocol~\cite{blum1983coin}, for example),
        server $A$ computes a check value
        ${c_A \gets \sigma \oplus \rho}$ and sends $c_A$ to the auditor.

  \item Server $B$ repeats Step~2 with ${\bf v}_B$.

  \item \label{step:clientsend}
        Since the client knows $\vA$, $\vB$, and $\sigma$, it can 
        compute $\mA$ and $\mB$ on its own.
        The client computes digests 
        $d_A = H({\bf m}_A)$ and $d_B = H({\bf m}_B)$,
        and sends these digests to the audit server.

  \item The audit server returns ``1'' to servers $A$ and $B$
        if and only if:
        \begin{itemize}
          \item the vectors it receives from the two
                servers are equal at every index except one, 
          \item the values $c_A$ and $c_B$ are equal, and
          \item the vectors ${\bf m}_A$ and ${\bf m}_B$ satisfy
                $d_A = H({\bf m}_A)$ and $d_B = H({\bf m}_B)$,
                where $d_A$ and $d_B$ are the client-provided digests. 
        \end{itemize}
        The auditor returns ``0'' otherwise.
        
\end{enumerate}

We include proofs of soundness, correctness, and zero-knowledge
for this construction in Appendix~\ref{app:almostequal}.

\medskip

The keys for the $(2,1)$-DPF construction 
have the form 
\[ k_A = ({\bf b}_A, {\bf s}_A, {\bf v}) \qquad
  k_B = ({\bf b}_B, {\bf s}_B, {\bf v}).\]
In a correctly formed pair of keys, the ${\bf b}$
and ${\bf s}$ vectors differ at a single index $\ell_x$,
and the ${\bf v}$ vector is equal to 
${\bf v} = m \cdot e_{\ell_y} + G({\bf s}_A[\ell_x]) + G({\bf s}_B[\ell_x])$.

To determine whether a pair of keys is correct,
server $A$ constructs a test vector ${\bf t}_A$
such that ${\bf t}_A[i] = {\bf b}_A[i] \| {\bf s}_A[i]$
for $i \in \{0, \dots, x-1\}$.
(where $\|$ denotes concatenation).
Server $B$ constructs a test vector ${\bf t}_B$
in the same way and the two servers, along with
the client and 
the auditor, run the protocol ${\sf AlmostEqual}({\bf t}_A, {\bf t}_B)$.
If the output of this protocol is ``1,'' then the servers
conclude that their ${\bf b}$ and ${\bf s}$ vectors
differ at a single index, though the protocol
does not reveal to the servers which index this is.
Otherwise, the servers reject the write request.

Next, the servers must verify that the ${\bf v}$ vector
is well-formed.
To do so, the servers compute another pair of test vectors:
\[ {\bf u}_A = \sum_{i=0}^{x-1} G({\bf s}_A[i]) \qquad
{\bf u}_B = {\bf v} + \sum_{i=0}^{x-1} G({\bf s}_B[i]). \]
The client and servers run ${\sf AlmostEqual}({\bf u}_A, {\bf u}_B)$
and accept the write request as valid if it returns ``1.''

We prove security of this construction in 
\abbr{the full version of this paper}{Appendix~\ref{app:smc-proof}}.

An important implementation note is that if $m = 0$---that is, if the
client writes the string of all zeros into the database---then 
the ${\bf u}$ vectors will not differ at any index and this information
is leaked to the auditor.
The protocol only provides security if the vectors differ at {\em exactly one}
index.
To avoid this information leakage, client requests 
must be defined such that $m \neq 0$ in every write request.
To achieve this, clients could define some special non-zero value
to indicate ``zero'' or could use a padding scheme to ensure
that zero values occur with negligible probability.

As a practical matter, the audit server needs to be able to match up the
portions of write requests coming from server $A$ with those coming from server
$B$. \Name achieves this as follows: 
When the client sends its upload request to server $A$, the client includes a
cryptographic hash of the request it sent to server $B$ (and vice versa). 
Both servers can use these hashes to derive a common nonce for the request. 
When the servers send audit requests to the audit server, they include the nonce
for the write request in question. 
The audit server can use the 
nonce to match every audit request from server $A$ with the corresponding request
from server $B$. 

\medskip

This three-party protocol is very efficient---it only 
requires $O(\sqrt{L})$ applications of a hash function and
$O(\sqrt{L})$ communication from the servers to the auditor.
The auditor only performs a simple string comparison, so it 
needs minimal computational and storage capabilities.

\subsection{Zero Knowledge Techniques}
\label{sec:disrupt:zkp}

Our second technique for detecting disruptors makes use
of non-interactive zero-knowledge proofs~\cite{camenisch1997proof,goldwasser1989knowledge,rackoff1992non}.

We apply zero-knowledge techniques to allow clients to prove the
well-formedness of their write requests.
This technique works in combination with the 
$(s,s-1)$-DPF presented in Section~\ref{sec:dpf:manyserver}
and maintains client write-privacy 
when {\em all but one} of $s$ servers is dishonest.

The keys for the $(s,s-1)$-DPF scheme are tuples 
$({\bf b}_i, {\bf s}_i, {\bf v})$ such that:
\[ \sum_{i=0}^{s-1} {\bf b}_i = e_{\ell_x} \qquad \sum_{i=0}^{s-1} {\bf s}_i = s^* \cdot e_{\ell_x} \qquad {\bf v} = m \cdot e_{\ell_y} -G(s^*)\]

To prove that its write request was correctly formed,
we have the client perform zero-knowledge proofs over
collections of Pedersen commitments~\cite{pedersen1992non}.
The public parameters for the Pedersen commitment scheme
consist of a group $\G$ of prime order $q$ and two generators
$P$ and $Q$ of $\G$ such that no one knows the discrete logarithm
$\log_Q P$.
A Pedersen commitment to a message $m \in \Z_q$ with randomness
$r \in \Z_q$ is $C(m,r) = (m P + r Q) \in \G$
(writing the group operation additively).
Pedersen commitments are {\em homomorphic}, in that given commitments
to $m_0$ and $m_1$, it is possible to compute a commitment to $m_0+m_1$:
\[ C(m_0,r_0) + C(m_1,r_1) = C(m_0+m_1,r_0+r_1) \]

Here, we assume that the $(s,s-1)$-DPF is instantiated with the
DDH-based PRG introduced in Section~\ref{sec:dpf:manyserver} and that
the group $\G$ used for the Pedersen commitments is the same order-$q$ group 
used in the PRG construction.

To execute the proof, the client first generates
Pedersen commitments to elements of each of the $s$ DPF keys.
Then each server $i$ can verify that the client computed the commitment
to the $i$-th DPF key elements correctly.
The servers use the homomorphic property of Pedersen commitments 
to generate commitments to the \textit{sum} of the elements of the DPF keys.
Finally, the client proves in zero knowledge that these sums have the 
correct values.

The protocols proceed as follows:

\begin{enumerate}
  \item The client generates vectors of Pedersen commitments ${\bf B}_i$
    and ${\bf S}_i$ committing to each element of ${\bf b}_i$ and ${\bf s}_i$.
    The client sends the ${\bf B}$ and ${\bf S}$ vectors to {\em every server}.
  \item To server $i$, the client sends the opening of the commitments
    ${\bf B}_i$ and ${\bf S}_i$.
    Each server $i$ verifies that ${\bf B}_i$ and ${\bf S}_i$ are valid
    commitments to the ${\bf b}_i$ and ${\bf s}_i$ vectors in the DPF key.
    If this check fails at some server $i$, server $i$
    notifies the other servers and all servers reject the write request.
  \item \label{item:check-commit}
    Using the homomorphic property of the commitments, each server can compute
    vectors of commitments ${\bf B}_\textrm{sum}$ and ${\bf S}_\textrm{sum}$
    to the vectors $\Sigma_{i=0}^{s-1} {\bf b}_i$ and $\Sigma_{i=0}^{s-1} {\bf s}_i$.
  \item \label{item:check-proof}
    Using a non-interactive zero-knowledge proof, the client proves to 
    the servers that ${\bf B}_\textrm{sum}$ and ${\bf S}_\textrm{sum}$
    are commitments to zero everywhere except at a single (secret) index $\ell_x$,
    and that ${\bf B}_\textrm{sum}[\ell_x]$ is a commitment to one.\footnote{
Technically, this is a zero-knowledge proof of knowledge
which proves that the client {\em knows} an opening of
the commitments to the stated values.
}
    This proof uses standard witness hiding techniques
    for discrete-logarithm-based zero knowledge 
    proofs~\cite{camenisch1997proof,cramer1994proofs}.
    If the proof is valid, the servers continue to check the ${\bf v}$ vector.
\end{enumerate}

This first protocol convinces each server that the ${\bf b}$ and ${\bf s}$ components
of the DPF keys are well formed.
Next, the servers check the ${\bf v}$ component:

\begin{enumerate}
  \item For each server $i$, the client sums up the seed values ${\bf s}_i$ it
        sent to server $i$: $\sigma_i = \Sigma_{j=0}^{s-1} {\bf s}_i[j]$.
        The client then generates the output of $G(\sigma_k)$ and blinds it: 
        \[{\bf G}_i = ({\sigma_i}P_1 + {r_1}Q,\, {\sigma_i}P_2 + r_2Q,\, \dots).\]
  \item The client sends the ${\bf G}$ values to all servers and the client
        sends the opening of ${\bf G}_i$ to each server $i$.
  \item Each server verifies that the openings are correct, and all servers
        reject the write request if this check fails at any server.
  \item Using the homomorphic property of Pedersen commitments, every server
        can compute a vector of commitments 
        ${\bf G}_\textrm{sum} = (\Sigma_{i=0}^{s-1} {\bf G}_i) + {\bf v}$.
        If ${\bf v}$ is well formed, then the ${\bf G}_\textrm{sum}$ vector
        contain commitments to zero at every index except one (at which it 
        will contain a commitment to the client's message $m$).
  \item \label{item:check-zkp}
        The client uses a non-interactive zero-knowledge proof to convince
        the servers that the vector of commitments
        ${\bf G}_\textrm{sum}$ contains commitments to zero at all indexes 
        except one.
        If the proof is valid, the servers accept the write request.
\end{enumerate}

We prove in \abbr{the full version of this paper}{Appendix~\ref{app:zkp-proof}}
that this protocol satisfies the
standard notions of soundness, completeness, and
zero-knowledge~\cite{camenisch1998group}.

 \section{Experimental Evaluation}
\label{sec:eval}

\begin{framed}
\noindent
{\color{blue} \textbf{Note:}}
As described on the title page, this is the extended and corrected
version of a paper by the same name that appeared at the
{\em IEEE Symposium on Security and Privacy}
in May 2015. This version corrects an error in the 
${\sf AlmostEqual}$ protocol of Section~\ref{sec:disrupt:smc}
that could allow a malicious database server to de-anonymize
a client using an active attack.

We thank Elette Boyle and Yuval Ishai for pointing out this
error and for helpful discussions on how to correct it.
Since the complexity of the updated protocol is almost identical
to the original one, and since the first author's 
dissertation~\cite[Section 5]{thesis} describes and evaluates
a DPF-checking protocol that subsumes the one
presented here, this evaluation section 
reflects the DPF-checking protocol from
the original version of this work.

The only important qualitative difference between the original
protocol and the one presented here is that the corrected protocol
models the hash function as a random oracle, and thus requires
using a strong hash function, such as SHA-256.
In contrast, the original protocol required a less expensive 
universal hash function, such as Poly1305.

When using Riposte with table size $L$,
we expect that the cost of the $O(L)$ AES operations required 
for DPF key expansion to dominate the $O(\sqrt L)$ hashing
operations needed for auditing.
Even so, we want to highlight this difference in case it
is relevant to those building on our work.
\end{framed}

To demonstrate that \name is a practical platform
for traffic-analysis-resistant anonymous messaging,
we implemented two variants of the system. 
The first variant uses the two-server distributed
point function (Section~\ref{sec:dpf:twoserver})
and uses the three-party protocol 
(Section~\ref{sec:disrupt:smc})
to prevent malicious clients from corrupting the database.
This variant is relatively fast, since it relies
primarily on symmetric-key primitives, but requires
that no two of the three servers collude.
Our results for the first variant {\em include the cost}
of identifying and excluding malicious clients.

The second variant uses the $s$-server distributed
point function (Section~\ref{sec:dpf:manyserver}).
This variant protects against $s-1$ colluding servers,
but relies on expensive public-key operations.
We have not implemented the zero-knowledge proofs
necessary to prevent disruptors for the $s$-server
protocol (Section~\ref{sec:disrupt:zkp}), so the
performance numbers represent only an upper bound on the
system throughput.

We wrote the prototype in the Go programming language
and have published the source code online at
\url{https://bitbucket.org/henrycg/riposte/}.
We used the DeterLab network 
testbed for our experiments~\cite{mirkovic2012teaching}.
All of the experiments used commodity servers running Ubuntu 14.04
with four-core AES-NI-enabled Intel E3-1260L CPUs 
and 16 GB of RAM.

Our experimental network topology used between
two and ten servers (depending on the protocol variant in use)
and eight client nodes.
In each of these experiments, the eight client machines
used many threads of execution to submit
write requests to the servers as quickly as possible.
In all experiments, the server nodes
connected to a common switch via 100 Mbps links,
the clients nodes connected to a common switch via 1 Gbps links,
and the client and server switches connected via a 1 Gbps link.
The round-trip network latency between each pair of nodes was 20~ms.
We chose this network topology to limit
the bandwidth between the servers to that of a fast WAN, but
to leave client bandwidth unlimited so that the small number
of client machines could saturate the servers with
write requests.

Error bars in the charts indicate the standard
deviation of the throughput measurements.

\subsection{Three-Server Protocol}

A three-server \name cluster consists of
two database servers and one audit server.
The system maintains its security properties as long
as {\em no two} of these three servers collude.
We have fully implemented the three-server protocol,
including the audit protocol (Section~\ref{sec:disrupt:smc}),
so the throughput numbers listed here {\em include} 
the cost of detecting and rejecting malicious write requests.

The prototype used AES-128 in counter mode as the pseudo-random
generator, Poly1305 as the keyed hash function 
used in the audit protocol~\cite{bernstein2005poly},
and TLS for link encryption.

\begin{figure}
\centering
\includegraphics[width=0.5\textwidth]{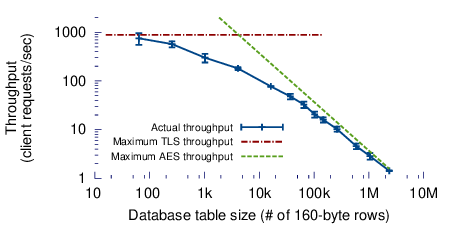}
\caption{As the database table size grows, the throughput of
  our system approaches the maximum possible given the 
  AES throughput of our servers.}
\label{fig:tablesize}
\end{figure}

Figure~\ref{fig:tablesize} shows how many client write requests
our \name cluster can service per second as the number of 160-byte
rows in the database table grows.
For a database table of 64 rows, the system handles 
751.5 write requests per second.
At a table size of 65,536 rows, the system handles 32.8 requests
per second.
At a table size of 1,048,576 rows, the system handles 2.86 requests
per second.

We chose the row length of 160 bytes because it was the smallest
multiple of 32 bytes large enough to to contain a 140-byte Tweet.
Throughput of the system depends only the total
size of the table (number of rows $\times$ row length), so larger
row lengths might be preferable for other applications.
For example, an anonymous email system using \name with 
4096-byte rows could handle 2.86 requests per second at a table
size of 40,960 rows.

An upper bound on the performance of the system is the speed 
of the pseudo-random generator used to stretch out the
DPF keys to the length of the database table.
The dashed line in Figure~\ref{fig:tablesize} indicates this upper
bound (605 MB/s), as determined using an AES benchmark written in Go.
That line indicates the maximum possible throughput we could hope
to achieve without aggressive optimization (e.g., writing portions of
the code in assembly) or more powerful machines.
Migrating the performance-critical
portions of our implementation from Go to C (using OpenSSL)
might increase the throughput by a factor of as much as
$6\times$, since ${\tt openssl\ speed}$ reports AES throughput of 3.9 GB/s,
compared with the 605 MB/s we obtain with Go's crypto library.
At very small table sizes, the speed at which the server can set up
TLS connections with the clients limits the overall throughput
to roughly 900 requests per second.

\begin{figure}
\centering
\includegraphics[width=0.5\textwidth]{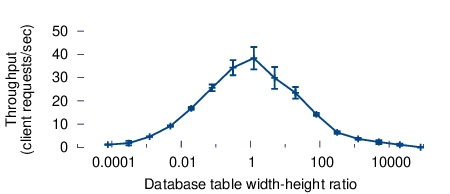}
\caption{Use of bandwidth-efficient DPFs gives 
 a $768\times$ speed-up over the na\"ive constructions,
 in which a client's request is as large as the database.}
\label{fig:tablewidth}
\end{figure}

Figure~\ref{fig:tablewidth} demonstrates how the request throughput
varies as the width of the table changes, while the number
of bytes in the table is held constant at 10 MB.
This figure demonstrates the performance advantage of using
a bandwidth-efficient $O(\sqrt{L})$ DPF (Section~\ref{sec:dpf})
over the na\"ive DPF (Section~\ref{sec:arch:straw}).
Using a DPF with optimal table size yields a throughput of 38.4 requests
per second.
The extreme left and right ends of the figure indicate the performance
yielded by the na\"ive construction, in which making a write request
involves sending a $(1 \times L)$-dimension vector to each server.
At the far right extreme of the table, performance drops to 
0.05 requests per second, so DPFs yield a 768$\times$ speed-up.

\begin{figure}
\centering
\includegraphics[width=0.5\textwidth]{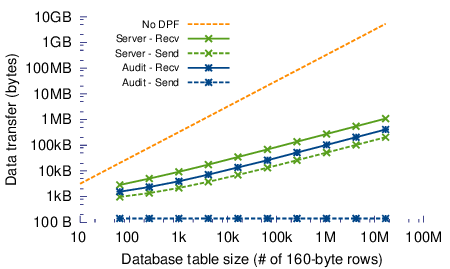}
\caption{The total client and server data 
  transfer scales sub-linearly
  with the size of the database.}
\label{fig:bandwidth}
\end{figure}

Figure~\ref{fig:bandwidth} indicates the total number of
bytes transferred by one of the database servers and
by the audit server while processing a single client write
request.
The dashed line at the top of the chart indicates the number
of bytes a client would need to send for a single write request
 if we did not use bandwidth-efficient DPFs
(i.e., the dashed line indicates the size of the database table).
As the figure demonstrates,
the total data transfer in a \name cluster
scales {\em sub-linearly} with the database size. 
When the database table is 2.5 GB in size, the database server
transfers only a total of 1.23 MB to process a write request.

\subsection{$s$-Server Protocol}

In some deployment scenarios, 
having strong protection against
server compromise may be more important
than performance or scalability.
In these cases, the $s$-server \name protocol 
provides the same basic functionality as the three-server
protocol described above, except that it maintains
privacy even if $s-1$ out of $s$ servers collude or
deviate arbitrarily from the protocol specification.
We implemented the basic $s$-server protocol but have not
yet implemented the zero-knowledge proofs necessary
to prevent malicious clients from corrupting the
database state (Section~\ref{sec:disrupt:zkp}).
These performance figures thus represent an {\em upper bound}
on the $s$-server protocol's performance.
Adding the zero-knowledge proofs 
would require an additional $\Theta(\sqrt{L})$
elliptic curve operations per server in an $L$-row database.
The computational cost of the proofs would almost certainly be dwarfed by the $\Theta(L)$
elliptic curve operations required to update the state of the database table.

The experiments use the DDH-based seed-homomorphic 
pseudo-random generator described
in Section~\ref{sec:dpf:manyserver} and they use the 
NIST P-256 elliptic curve as the underlying algebraic group.
The table row size is fixed at 160 bytes. 

\begin{figure}
\centering
\includegraphics[width=0.5\textwidth]{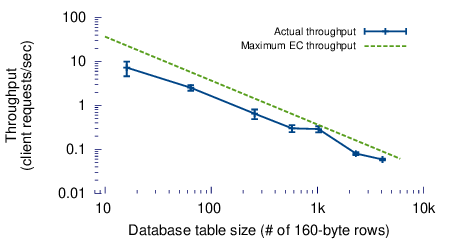}
\caption{Throughput of an eight-server \name cluster using the 
  $(8,7)$-distributed point function.}
\label{fig:ddh-tablesize}
\end{figure}

Figure~\ref{fig:ddh-tablesize} demonstrates the performance
of an eight-server \name cluster as the table size increases.
At a table size of 1,024 rows, the cluster can process
one request every 3.44 seconds.
The limiting factor is the rate at which the servers can evaluate
the DDH-based pseudo-random generator (PRG),
since computing each 32-byte block of PRG output requires 
a costly elliptic curve scalar multiplication.
The dashed line in the figure indicates the maximum throughput
obtainable using Go's implementation of P-256 on our servers,
which in turn dictates the maximum cluster throughput.
Processing a single request with a table size of one million
rows would take nearly one hour with this construction, compared
to 0.3 seconds in the AES-based three-server protocol.

\begin{figure}
\centering
\includegraphics[width=0.5\textwidth]{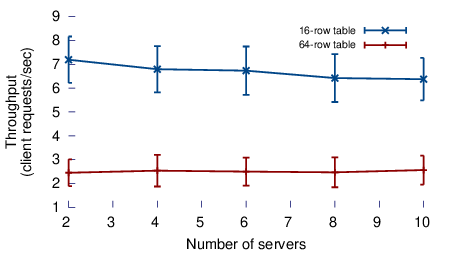}
\caption{Throughput of \name clusters using two
    different database table sizes as the number 
    of servers varies.}
\label{fig:ddh-servers}
\end{figure}

Figure~\ref{fig:ddh-servers} shows how the throughput of the
\name cluster changes as the number of servers varies.
Since the workload is heavily CPU-bound, the throughput
only decreases slightly as the number of servers increases
from two to ten.

\subsection{Discussion: Whistleblowing and\\ Microblogging with Million-User\\ Anonymity Sets}
\label{sec:eval:million}

Whistleblowers, political activists,
or others discussing sensitive or controversial
issues might benefit from an anonymous microblogging service.
A whistleblower, for example, might want to anonymously blog
about an instance of bureaucratic corruption in her organization.
The utility of such a system depends on the size of the anonymity set
it would provide:
if a whistleblower is only anonymous amongst a group of ten people,
it would be easy for the whistleblower's employer to retaliate against
{\em everyone} in the anonymity set. 
Mounting this ``punish-them-all'' attack does not require breaking 
the anonymity system itself, since the anonymity set is public.
As the anonymity set size grows, however, the
feasibility of the ``punish-them-all'' attack quickly tends to zero.
At an anonymity set size of 1,000,000 clients, mounting an 
``punish-them-all'' attack would be prohibitively expensive
in most situations.

\Name can handle such large anonymity sets as long as
(1) clients are willing to tolerate hours of messaging latency, and 
(2) only a small fraction of clients writes into the database in each time epoch.
Both of these requirements are satisfied in the whistleblowing
scenario.
First, whistleblowers might not care if the system delays their posts by
a few hours.
Second, the vast majority of users of a microblogging service
(especially in the whistleblowing context) are 
more likely to {\em read} posts than write them.
To get very large anonymity sets, maintainers of an anonymous
microblogging service could take advantage of the large set of
``read-only'' users to provide anonymity for the relatively 
small number of ``read-write'' users.

The client application for such a microblogging service would 
enable read-write users to generate and submit
\name write requests to a \name cluster running the microblogging service.
However, the client application would 
also allow read-only users to submit an ``empty''
write request to the \name cluster that would always write a random
message into the first row of the \name database.
From the perspective of the servers, a read-only client would be 
indistinguishable from a read-write client.
By leveraging read-only users in this way, 
we can increase the size of the anonymity set without
needing to increase the size of the database table.

To demonstrate that \name can support very large anonymity set
sizes---albeit with high latency---we configured a cluster
of \name servers with a 65,536-row database table and left it running for 32 hours.
In that period, the system processed a total of 2,895,216 
write requests at an average rate of 25.19 requests per second.
(To our knowledge, this is the largest
anonymity set {\em ever constructed} in a system that
offers protection against traffic analysis attacks.)
Using the techniques in Section~\ref{sec:collisions},
a table of this size could handle 0.3\% of users writing at
a collision rate of under 5\%.
Thus, to get an anonymity set of roughly 1,000,000 users
with a three-server \name cluster and
a database table of size $65,536$, the time epoch must 
be at least 11 hours long.

As of 2013, Twitter reported an average throughput of
5,700 140-byte Tweets per second~\cite{krikorian2013new}.
That is equivalent roughly 5,000 of our 160-byte messages per second.
At a table size of one million messages, our \name cluster's 
end-to-end throughput is 2.86 write requests per second (Figure~\ref{fig:tablesize}).
To handle the same volume of Tweets as Twitter does with
anonymity set sizes on the order of hundreds of thousands of clients, 
we would need to increase the computing power of our cluster by ``only''
a factor of $\approx$1,750.\footnote{
We assume here that scaling the number of machines
by a factor of $k$ increases our throughput by a factor of $k$.
This assumption is reasonable given our workload, since 
the processing of write requests is an embarrassingly parallel task.
}
Since we are using only three servers now, we would need
roughly 5,250 servers (split into three non-colluding data centers)
to handle the same volume of traffic as Twitter. 
Furthermore, since the audit server is just doing string comparisons,
the system would likely need many fewer audit servers than database servers,
so the total number of servers required might be closer to $4,000$.

 \section{Related Work}
\label{sec:rel}

Anonymity systems fall into one of two general categories: systems that provide
low-latency communication and those that protect against traffic analysis
attacks by a global network adversary.

Aqua~\cite{leblond2013towards}, 
Crowds~\cite{reiter1998crowds}, 
LAP~\cite{hsiao2012lap}, 
ShadowWalker~\cite{mittal2009shadowwalker},
Tarzan~\cite{freedman2002tarzan}, and
Tor~\cite{dingledine2004tor} 
belong to the first category of systems: 
they provide an anonymous proxy for 
real-time Web browsing, but they
do not protect against an adversary 
who controls the network, many of
the clients, and some of the nodes
on a victim's path through the network.
Even providing a formal 
definition of anonymity for low-latency systems
is challenging~\cite{johnson2009design} and such
definitions typically do not capture the need to 
protect against timing attacks.

Even so, it would be possible to combine 
Tor (or another low-latency anonymizing proxy) and \name
to build a ``best of both'' anonymity system: clients would submit
their write requests to the \name servers via the Tor network.
In this configuration, even if {\em all} of the \name servers
colluded, they could not learn which user wrote which message
without also breaking the anonymity of the Tor network.

David Chaum's ``cascade'' mix networks were one of the
first systems devised with the specific goal of defending
against traffic-analysis attacks~\cite{chaum1981untraceable}.
Since then, there have been a number of mix-net-style systems proposed, 
many of which explicitly weaken 
their protections against a near omni-present adversary~\cite{syverson2013why}
to improve prospects for practical usability 
(i.e., for email traffic)~\cite{danezis2003mixminion}.
In contrast, \Name attempts to provide very strong anonymity guarantees
at the price of usability for interactive applications.

E-voting systems (also called ``verifiable shuffles'')
achieve the sort of privacy properties that \name 
offers, and some systems even provide stronger 
voting-specific guarantees 
(receipt-freeness, proportionality, etc.),
though most e-voting systems cannot provide
the forward security property that \name offers (Section~\ref{sec:arch:forward})~\cite{adida2008helios,clarkson2007civitas,neff2001verifiable,groth2007verifiable,furukawa2004efficient,groth2010verifiable,rabin2014efficient}.

In a typical e-voting system, voters submit their encrypted ballots
to a few trustees, who collectively shuffle and decrypt them.
While it is possible to repurpose e-voting systems for anonymous messaging,
they typically require expensive zero-knowledge proofs 
or are inefficient when message sizes are large.
Mix-nets that do not use zero-knowledge proofs of correctness
typically do not provide privacy in the face of active attacks 
by a subset of the mix servers. 

For example, the verifiable shuffle protocol of 
Bayer and Groth~\cite{bayer2012efficient}
is one of the most efficient in the literature.
Their shuffle implementation, when used with 
an anonymity set of size $N$, requires $16N$ group 
exponentiations per server and data
transfer $O(N)$.
In addition, messages must be small enough to be encoded
in single group elements (a few hundred bytes at most).
In contrast, our protocol requires $O(L)$ AES operations and
data transfer $O(\sqrt{L})$, where $L$ is the size of the
database table.
When messages are short and when the writer/reader ratio
is high, the Bayer-Groth mix may be faster than our system.
In contrast, when messages are long and when the writer/reader
ratio is low (i.e., $L \ll O(N)$), our system is faster.

Chaum's Dining Cryptographers network
(DC-net) is an information-theoretically secure
anonymous broadcast channel~\cite{chaum1988dining}.
A DC-net provides the same strong anonymity properties
as \name does, but it requires every user of a DC-net
to participate in every run of the protocol.
As the number of users grows, this quickly becomes impractical.

The Dissent~\cite{wolinsky2012dissent} system
introduced the idea of using partially trusted servers to 
make DC-nets practical in distributed networks. 
Dissent requires weaker trust assumptions than our three-server
protocol does but it requires clients to send $O(L)$ bits
to each server per time epoch (compared with our $O(\sqrt{L})$).
Also, excluding {\em a single} disruptor in a 1,000-client
deployment takes over an hour.
In contrast, \name can excludes disruptors as fast as it processes write
requests (tens to hundreds per second, depending on the database size).
Recent work~\cite{corrigangibbs2013proactively} uses
zero-knowledge techniques to speed up disruption resistance
in Dissent (building on ideas of Golle and Juels~\cite{golle2004dining}).
Unfortunately, these techniques limit the system's end to end-throughput
end-to-end throughput to 30 KB/s, compared with \name's 450+ MB/s.

Herbivore scales DC-nets by dividing users into many small anonymity
sets~\cite{goel2003herbivore}.
\Name creates a single large anonymity set, and thus enables
every client to be anonymous amongst the {\em entire set} of honest clients.

Our DPF constructions make extensive use
of prior work on private information
retrieval (PIR)~\cite{chor1998private,chor1997computationally,gasarch2004survey,gilboa2014distributed}.
Recent work demonstrates that it is possible to make
theoretical PIR fast enough for practical 
use~\cite{devet2014best,goldberg2007improving,demmler2014raid}.
Function secret sharing~\cite{boyle2015function} generalizes DPFs
to allow sharing of more sophisticated functions (rather than just
point functions).
This more powerful primitive may prove useful for PIR and anonymous
messaging applications in the future.

Gertner et al.\cite{gertner1998protecting} consider {\em symmetric}
PIR protocols, in which the servers prevent dishonest clients
from learning about more than a single row of the database per query.
The problem that Gertner et al.\ consider is, in a way, the dual 
of the problem we address in Section~\ref{sec:disrupt}, though
their techniques do not appear to apply directly in our setting.

Ostrovsky and Shoup first proposed
using PIR protocol as the basis for writing
into a database shared across a set of servers~\cite{ostrovsky1997private}.
However, Ostrovsky and Shoup
considered only the case of a single honest client, 
who uses the untrusted database servers for private storage. 
Since {\em many mutually distrustful clients} use a
single \name cluster, our protocol must also handle malicious clients.

Pynchon Gate~\cite{sassaman2005pynchon} builds a 
private point-to-point messaging system from mix-nets and PIR.
Clients anonymously upload messages to email servers using a traditional mix-net
and download messages from the email servers using a PIR protocol.
\Name could replace the mix-nets used in the Pynchon Gate system: clients
could anonymously write their messages into the database using \name
and could privately read incoming messages using PIR.

 \section{Conclusion and Open Questions}
\label{sec:concl}

We have presented \name, a new system for anonymous messaging.
To the best of our knowledge, \name is the first system
that simultaneously 
(1) thwarts traffic analysis attacks,
(2) prevents malicious clients from anonymously disrupting the system, and
(3) enables million-client anonymity set sizes.
We achieve these goals through novel application of private information
retrieval and secure multiparty computation techniques.
We have demonstrated \name's practicality by implementing it and
evaluating it with anonymity sets of over two million nodes.
This work leaves open a number of questions for future work, including:
\begin{compactitem}
\item Does there exist an $(s,s-1)$-DPF construction for $s > 2$
      that uses only symmetric-key operations?
\item Are there efficient techniques 
      (i.e., using no public-key primitives) for achieving
      disruption resistance without the need for a
      non-colluding audit server?
\item Are there DPF constructions that enable processing
      write requests in amortized time $o(L)$, for a 
      length-$L$ database?
\end{compactitem}
With the design and implementation of \name,
we have demonstrated that cryptographic techniques can 
make traffic-analysis-resistant anonymous microblogging
and whistleblowing more practical at Internet scale.

\subsection*{Acknowledgements}
We thank Elette Boyle and Yuval Ishai for pointing out an error in 
the ${\sf AlmostEqual}$ protocol of Section~\ref{sec:disrupt:smc},
and for helpful discussions on how to repair it.
We thank Paul Grubbs for the suggestion to add a note on this bug
to the evaluation section of the paper.
We would like to thank Joe Zimmerman and David
Wu for helpful discussions about distributed point functions.
We would like to thank Stephen Schwab and the staff of
DeterLab for giving us access to their excellent network testbed. 
This work was supported by NSF, an IARPA project provided via DoI/NBC,
a grant from ONR, an NDSEG fellowship, 
and by a Google faculty scholarship. Opinions,
findings and conclusions or recommendations expressed in this material
are those of the author(s) and do not necessarily reflect the views of
DARPA or IARPA.

\frenchspacing
\bibliographystyle{plain}
\bibliography{refs}

\begin{thebibliography}{10}

\bibitem{adida2008helios}
Ben Adida.
\newblock Helios: {Web}-based open-audit voting.
\newblock In {\em USENIX Security Symposium}, volume~17, 2008.

\bibitem{adida2007shuffle}
Ben Adida and Douglas Wikstr{\"o}m.
\newblock How to shuffle in public.
\newblock In {\em Theory of Cryptography}. 2007.

\bibitem{banerjee2014new}
Abhishek Banerjee and Chris Peikert.
\newblock New and improved key-homomorphic pseudorandom functions.
\newblock In {\em CRYPTO}, 2014.

\bibitem{bauer2007low}
Kevin Bauer, Damon McCoy, Dirk Grunwald, Tadayoshi Kohno, and Douglas Sicker.
\newblock Low-resource routing attacks against {Tor}.
\newblock In {\em WPES}. ACM, 2007.

\bibitem{bayer2012efficient}
Stephanie Bayer and Jens Groth.
\newblock Efficient zero-knowledge argument for correctness of a shuffle.
\newblock In {\em EUROCRYPT}. 2012.

\bibitem{bellare1993random}
Mihir Bellare and Phillip Rogaway.
\newblock Random oracles are practical: A paradigm for designing efficient
  protocols.
\newblock In {\em CCS}. ACM, 1993.

\bibitem{bennhold2014britain}
Katrin Bennhold.
\newblock In {Britain}, guidelines for spying on lawyers and clients.
\newblock {\em New York Times}, page~A6, 7 Nov. 2014.

\bibitem{bernstein2005poly}
Daniel~J Bernstein.
\newblock The {Poly1305-AES} message-authentication code.
\newblock In {\em Fast Software Encryption}, 2005.

\bibitem{blum1983coin}
Manuel Blum.
\newblock Coin flipping by telephone: a protocol for solving impossible
  problems.
\newblock {\em ACM SIGACT News}, 15(1):23--27, 1983.

\bibitem{boneh1998decision}
Dan Boneh.
\newblock The decision {Diffie-Hellman} problem.
\newblock In Joe~P. Buhler, editor, {\em Algorithmic Number Theory}, volume
  1423 of {\em Lecture Notes in Computer Science}, pages 48--63. Springer,
  1998.

\bibitem{boneh2013key}
Dan Boneh, Kevin Lewi, Hart Montgomery, and Ananth Raghunathan.
\newblock Key homomorphic {PRFs} and their applications.
\newblock In {\em CRYPTO}. 2013.

\bibitem{bos1989detection}
Jurjen Bos and Bert den Boer.
\newblock Detection of disrupters in the {DC} protocol.
\newblock In {\em EUROCRYPT}, 1989.

\bibitem{boyle2015function}
Elette Boyle, Niv Gilboa, and Yuval Ishai.
\newblock Function secret sharing.
\newblock In {\em EUROCRYPT}, pages 337--367, 2015.

\bibitem{boyle2016function}
Elette Boyle, Niv Gilboa, and Yuval Ishai.
\newblock Function secret sharing: Improvements and extensions.
\newblock In {\em CCS}, pages 1292--1303. ACM, 2016.

\bibitem{camenisch1997proof}
Jan Camenisch and Markus Stadler.
\newblock Proof systems for general statements about discrete logarithms.
\newblock Technical Report 260, Dept. of Computer Science, ETH Zurich, March
  1997.

\bibitem{camenisch1998group}
Jan~Leonhard Camenisch.
\newblock {\em Group Signature Schemes and Payment Systems Based on the
  Discrete Logarithm Problem}.
\newblock PhD thesis, Swiss Federal Institute of Technology Z{\"u}rich (ETH
  Z{\"u}rich), 1998.

\bibitem{canetti2003forward}
Ran Canetti, Shai Halevi, and Jonathan Katz.
\newblock A forward-secure public-key encryption scheme.
\newblock In {\em EUROCRYPT}. 2003.

\bibitem{chaum1988dining}
David Chaum.
\newblock The {Dining Cryptographers} problem: Unconditional sender and
  recipient untraceability.
\newblock {\em Journal of Cryptology}, pages 65--75, January 1988.

\bibitem{chaum1981untraceable}
David~L Chaum.
\newblock Untraceable electronic mail, return addresses, and digital
  pseudonyms.
\newblock {\em Communications of the ACM}, 24(2):84--90, 1981.

\bibitem{chien1966application}
Robert~T Chien and WD~Frazer.
\newblock An application of coding theory to document retrieval.
\newblock {\em Information Theory, IEEE Transactions on}, 12(2):92--96, 1966.

\bibitem{chor1997computationally}
Benny Chor and Niv Gilboa.
\newblock Computationally private information retrieval.
\newblock In {\em STOC}. ACM, 1997.

\bibitem{chor1998private}
Benny Chor, Eyal Kushilevitz, Oded Goldreich, and Madhu Sudan.
\newblock Private information retrieval.
\newblock {\em Journal of the ACM}, 45(6):965--981, 1998.

\bibitem{clarkson2007civitas}
Michael~R Clarkson, Stephen Chong, and Andrew~C Myers.
\newblock Civitas: A secure voting system.
\newblock Technical Report TR 2007-2081, Cornell University, May 2007.

\bibitem{thesis}
Henry Corrigan-Gibbs.
\newblock {\em Protecting Privacy by Splitting Trust}.
\newblock PhD thesis, Stanford University, December 2019.

\bibitem{corrigangibbs2010dissent}
Henry Corrigan-Gibbs and Bryan Ford.
\newblock {Dissent}: Accountable anonymous group messaging.
\newblock In {\em CCS}. ACM, October 2010.

\bibitem{corrigangibbs2013proactively}
Henry Corrigan-Gibbs, David~Isaac Wolinsky, and Bryan Ford.
\newblock Proactively accountable anonymous messaging in {Verdict}.
\newblock In {\em USENIX Security Symposium}, 2013.

\bibitem{cramer1994proofs}
Ronald Cramer, Ivan Damg{\aa}rd, and Berry Schoenmakers.
\newblock Proofs of partial knowledge and simplified design of witness hiding
  protocols.
\newblock In {\em CRYPTO}, 1994.

\bibitem{danezis2008survey}
George Danezis and Claudia Diaz.
\newblock A survey of anonymous communication channels.
\newblock Technical report, Technical Report MSR-TR-2008-35, Microsoft
  Research, 2008.

\bibitem{danezis2003mixminion}
George Danezis, Roger Dingledine, and Nick Mathewson.
\newblock {Mixminion}: Design of a type {III} anonymous remailer protocol.
\newblock In {\em Security and Privacy}. IEEE, 2003.

\bibitem{danezis2004statistical}
George Danezis and Andrei Serjantov.
\newblock Statistical disclosure or intersection attacks on anonymity systems.
\newblock In {\em Information Hiding Workshop}, May 2004.

\bibitem{demmler2014raid}
Daniel Demmler, Amir Herzberg, and Thomas Schneider.
\newblock {RAID-PIR}: Practical multi-server {PIR}.
\newblock In {\em WPES}, 2014.

\bibitem{devet2014best}
Casey Devet and Ian Goldberg.
\newblock The best of both worlds: Combining information-theoretic and
  computational pir for communication efficiency.
\newblock In {\em PETS}, July 2014.

\bibitem{dingledine2004tor}
Roger Dingledine, Nick Mathewson, and Paul Syverson.
\newblock {Tor}: The second-generation onion router.
\newblock In {\em USENIX Security Symposium}, August 2004.

\bibitem{edman2009anonymity}
Matthew Edman and B{\"u}lent Yener.
\newblock On anonymity in an electronic society: A survey of anonymous
  communication systems.
\newblock {\em ACM Computing Surveys}, 42(1):5, 2009.

\bibitem{fagin1996comparing}
Ronald Fagin, Moni Naor, and Peter Winkler.
\newblock Comparing information without leaking it.
\newblock {\em Communications of the ACM}, 39(5):77--85, 1996.

\bibitem{feige1988zero}
Uriel Feige, Amos Fiat, and Adi Shamir.
\newblock Zero-knowledge proofs of identity.
\newblock {\em Journal of Cryptology}, 1(2):77--94, 1988.

\bibitem{freedman2002tarzan}
Michael~J Freedman and Robert Morris.
\newblock Tarzan: A peer-to-peer anonymizing network layer.
\newblock In {\em CCS}. ACM, 2002.

\bibitem{furukawa2004efficient}
Jun Furukawa.
\newblock Efficient, verifiable shuffle decryption and its requirement of
  unlinkability.
\newblock In {\em PKC}. 2004.

\bibitem{gasarch2004survey}
William Gasarch.
\newblock A survey on private information retrieval.
\newblock In {\em Bulletin of the EATCS}, 2004.

\bibitem{gellman2013nsa}
Barton Gellman and Ashkan Soltani.
\newblock {NSA} infiltrates links to {Yahoo}, {Google} data centers worldwide,
  {Snowden} documents say.
\newblock {\em Washington Post}, Oct. 30 2013.

\bibitem{gellman2014nsa}
Barton Gellman, Julie Tate, and Ashkan Soltani.
\newblock In {NSA}-intercepted data, those not targeted far outnumber the
  foreigners who are.
\newblock {\em Washington Post}, 5 Jul. 2014.

\bibitem{gertner1998protecting}
Yael Gertner, Yuval Ishai, Eyal Kushilevitz, and Tal Malkin.
\newblock Protecting data privacy in private information retrieval schemes.
\newblock In {\em STOC}, 1998.

\bibitem{gilboa2014distributed}
Niv Gilboa and Yuval Ishai.
\newblock Distributed point functions and their applications.
\newblock In {\em EUROCRYPT}. 2014.

\bibitem{goel2003herbivore}
Sharad Goel, Mark Robson, Milo Polte, and Emin Sirer.
\newblock Herbivore: A scalable and efficient protocol for anonymous
  communication.
\newblock Technical report, Cornell University, 2003.

\bibitem{goel2014government}
Vindu Goel.
\newblock Government push for {Yahoo}'s user data set stage for broad
  surveillance.
\newblock {\em New York Times}, page~B3, 7 Sept. 2014.

\bibitem{goldberg2007improving}
Ian Goldberg.
\newblock Improving the robustness of private information retrieval.
\newblock In {\em Security and Privacy}. IEEE, 2007.

\bibitem{goldreich1987play}
Oded Goldreich, Silvio Micali, and Avi Wigderson.
\newblock How to play any mental game.
\newblock In {\em STOC}. ACM, 1987.

\bibitem{goldreich1991proofs}
Oded Goldreich, Silvio Micali, and Avi Wigderson.
\newblock Proofs that yield nothing but their validity or all languages in {NP}
  have zero-knowledge proof systems.
\newblock {\em Journal of the ACM}, 38(3):690--728, 1991.

\bibitem{goldwasser1989knowledge}
Shafi Goldwasser, Silvio Micali, and Charles Rackoff.
\newblock The knowledge complexity of interactive proof systems.
\newblock {\em SIAM Journal on computing}, 18(1):186--208, 1989.

\bibitem{golle2004dining}
Philippe Golle and Ari Juels.
\newblock Dining cryptographers revisited.
\newblock In {\em EUROCRYPT}, 2004.

\bibitem{groth2010verifiable}
Jens Groth.
\newblock A verifiable secret shuffle of homomorphic encryptions.
\newblock {\em Journal of Cryptology}, 23(4):546--579, 2010.

\bibitem{groth2007verifiable}
Jens Groth and Steve Lu.
\newblock Verifiable shuffle of large size ciphertexts.
\newblock In {\em PKC}. 2007.

\bibitem{haastad1999pseudorandom}
Johan H{\aa}stad, Russell Impagliazzo, Leonid~A Levin, and Michael Luby.
\newblock A pseudorandom generator from any one-way function.
\newblock {\em SIAM Journal on Computing}, 28(4):1364--1396, 1999.

\bibitem{hsiao2012lap}
Hsu-Chun Hsiao, TH-J Kim, Adrian Perrig, Akira Yamada, Samuel~C Nelson, Marco
  Gruteser, and Wei Meng.
\newblock {LAP}: Lightweight anonymity and privacy.
\newblock In {\em Security and Privacy}. IEEE, May 2012.

\bibitem{johnson2009design}
Aaron Johnson.
\newblock {\em Design and Analysis of Efficient Anonymous-Communication
  Protocols}.
\newblock PhD thesis, Yale University, December 2009.

\bibitem{rfc7296}
C~Kaufman, P~Hoffman, Y~Nir, P~Eronen, and Kivinen T.
\newblock {RFC7296}: Internet key exchange protocol version 2 ({IKEv2}),
  October 2014.

\bibitem{kedogan2003limits}
Dogan Kedogan, Dakshi Agrawal, and Stefan Penz.
\newblock Limits of anonymity in open environments.
\newblock In {\em Information Hiding}, 2003.

\bibitem{krikorian2013new}
Raffi Krikorian.
\newblock New {Tweets} per second record, and how!
\newblock
  \url{https://blog.twitter.com/2013/new-tweets-per-second-record-and-how},
  August 2013.

\bibitem{leblond2013towards}
Stevens Le~Blond, David Choffnes, Wenxuan Zhou, Peter Druschel, Hitesh Ballani,
  and Paul Francis.
\newblock Towards efficient traffic-analysis resistant anonymity networks.
\newblock In {\em SIGCOMM}. ACM, 2013.

\bibitem{liskov2012viewstamped}
Barbara Liskov and James Cowling.
\newblock Viewstamped replication revisited.
\newblock Technical Report MIT-CSAIL-TR-2012-021, MIT CSAIL, July 2013.

\bibitem{mathewson2005practical}
Nick Mathewson and Roger Dingledine.
\newblock Practical traffic analysis: Extending and resisting statistical
  disclosure.
\newblock In {\em Privacy Enhancing Technologies}, 2005.

\bibitem{miller1986use}
Victor~S Miller.
\newblock Use of elliptic curves in cryptography.
\newblock In {\em CRYPTO}, 1986.

\bibitem{mirkovic2012teaching}
Jelena Mirkovic and Terry Benzel.
\newblock Teaching cybersecurity with {DeterLab}.
\newblock {\em Security \& Privacy}, 10(1), 2012.

\bibitem{mittal2009shadowwalker}
Prateek Mittal and Nikita Borisov.
\newblock {ShadowWalker}: Peer-to-peer anonymous communication using redundant
  structured topologies.
\newblock In {\em CCS}. ACM, November 2009.

\bibitem{murdoch2005low}
Steven~J Murdoch and George Danezis.
\newblock Low-cost traffic analysis of {Tor}.
\newblock In {\em Security and Privacy}. IEEE, 2005.

\bibitem{murdoch2007sampled}
Steven~J. Murdoch and Piotr Zieli{\'n}ski.
\newblock Sampled traffic analysis by {I}nternet-exchange-level adversaries.
\newblock In {\em PETS}, June 2007.

\bibitem{nakashima2014court}
Ellen Nakashima and Barton Gellman.
\newblock Court gave {NSA} broad leeway in surveillance, documents show.
\newblock {\em Washington Post}, 30 Jun. 2014.

\bibitem{naor1999distributed}
Moni Naor, Benny Pinkas, and Omer Reingold.
\newblock Distributed pseudo-random functions and {KDC}s.
\newblock In {\em EUROCRYPT}, 1999.

\bibitem{nist2001aes}
{National Institute of Standards and Technology}.
\newblock Specification for the advanced encryption standard ({AES}).
\newblock Federal Information Processing Standards Publication 197, November
  2001.

\bibitem{neff2001verifiable}
C~Andrew Neff.
\newblock A verifiable secret shuffle and its application to e-voting.
\newblock In {\em CCS}. ACM, 2001.

\bibitem{ongaro2014search}
Diego Ongaro and John Ousterhout.
\newblock In search of an understandable consensus algorithm.
\newblock In {\em ATC}. USENIX, June 2014.

\bibitem{ostrovsky1997private}
Rafail Ostrovsky and Victor Shoup.
\newblock Private information storage.
\newblock In {\em STOC}, 1997.

\bibitem{pedersen1992non}
Torben~Pryds Pedersen.
\newblock Non-interactive and information-theoretic secure verifiable secret
  sharing.
\newblock In {\em CRYPTO}, 1992.

\bibitem{rabin2014efficient}
Michael~O. Rabin and Ronald~L. Rivest.
\newblock Efficient end to end verifiable electronic voting employing split
  value representations.
\newblock In {\em EVOTE 2014}, August 2014.

\bibitem{rackoff1992non}
Charles Rackoff and Daniel~R Simon.
\newblock Non-interactive zero-knowledge proof of knowledge and chosen
  ciphertext attack.
\newblock In {\em CRYPTO}, 1992.

\bibitem{reiter1998crowds}
Michael~K Reiter and Aviel~D Rubin.
\newblock {Crowds}: Anonymity for {Web} transactions.
\newblock {\em ACM Transactions on Information and System Security},
  1(1):66--92, 1998.

\bibitem{sassaman2005pynchon}
Len Sassaman, Bram Cohen, and Nick Mathewson.
\newblock The {Pynchon} gate: A secure method of pseudonymous mail retrieval.
\newblock In {\em WPES}, November 2005.

\bibitem{serjantov2003trickle}
Andrei Serjantov, Roger Dingledine, and Paul Syverson.
\newblock From a trickle to a flood: Active attacks on several mix types.
\newblock In {\em Information Hiding}, 2003.

\bibitem{syverson2013why}
Paul Syverson.
\newblock Why {I}'m not an entropist.
\newblock In {\em Security Protocols XVII}. 2013.

\bibitem{waidner1989dining}
Michael Waidner and Birgit Pfitzmann.
\newblock The {Dining Cryptographers} in the disco: Unconditional sender and
  recipient untraceability with computationally secure serviceability.
\newblock In {\em EUROCRYPT}, April 1989.

\bibitem{wolinsky2013hang}
David Wolinsky, Ewa Syta, and Bryan Ford.
\newblock Hang with your buddies to resist intersection attacks.
\newblock In {\em CCS}, November 2013.

\bibitem{wolinsky2012dissent}
David~Isaac Wolinsky, Henry Corrigan-Gibbs, Aaron Johnson, and Bryan Ford.
\newblock Dissent in numbers: Making strong anonymity scale.
\newblock In {\em 10th OSDI}. USENIX, October 2012.

\bibitem{yao1982protocols}
Andrew~C Yao.
\newblock Protocols for secure computations.
\newblock In {\em FOCS}. IEEE, 1982.

\end{thebibliography}
\nonfrenchspacing

\appendix
\section{Definition of Write Privacy}
\label{app:game}

An $(s,t)$-{\em write-private database scheme}
consists of the following three (possibly randomized) algorithms: 

\medskip

\begin{hangparas}{0.25in}{1}
  ${\sf Write}(\ell, m) \to (w^{(0)}, \dots, w^{(s-1)})$.
  Clients use the ${\sf Write}$ functionality to generate the
  write request queries sent to the $s$ servers. 
  The ${\sf Write}$ function takes as input a message $m$ (from some 
  finite message space) and an integer $\ell$ and produces a set of $s$
  write requests---one per server.
  
  ${\sf Update}(\sigma, w) \to \sigma'$.
  Servers use the ${\sf Update}$ functionality to process incoming
  write requests.
  The ${\sf Update}$ function takes as input a server's internal state
  $\sigma$, a write request $w$, and outputs the updated state of the server $\sigma'$.

  ${\sf Reveal}(\sigma_0, \dots, \sigma_{s-1}) \to D$.
  At the end of the time epoch, servers use the ${\sf Reveal}$ functionality
  to recover the contents of the database.
  The ${\sf Reveal}$ function takes as input the set of states from
  each of the $s$ servers and produces the plaintext database contents $D$.
\end{hangparas}

\medskip

We define the write-privacy property using the following
security game, played between the adversary (who statically
corrupts up to $t$ servers and all but two clients) and a challenger.
\begin{enumerate}
  \item In the first step, the adversary performs the following actions: 
        \begin{itemize}
            \item The adversary selects 
              a subset $\Adv_s \subseteq \{0, \dots, s-1\}$
              of the servers, such that $|\Adv_s| \leq t$.
              The set $\Adv_s$ represents the
              set of adversarial servers. 
              Let the set $\HH_s = \{0, \dots, s-1\} \setminus \Adv_s$
              represent the set of honest servers.

            \item The adversary selects a set of clients $\HH_c \subseteq \{0, \dots, n-1\}$,
                  such that $|\HH_c| \geq 2$, representing the set of honest clients.
                  The adversary selects
                  one message-location pair per honest client:
                \[ \MM = \{ (i, m_i, \ell_i) \ |\ i \in \HH_c \} \]

        \end{itemize}
        The adversary sends $\Adv_s$ and $\MM$ to the challenger.

  \item In the second step, the challenger responds to the adversary:
        \begin{itemize}
        \item 
        For each $(i, m_i, \ell_i) \in \mathcal{M}$, the challenger generates 
        a write request: 
        \[ (w^{(0)}_i, \dots, w^{(s-1)}_{i}) \gets {\sf Write}(\ell_i, m_i) \]
        The set of shares of the $i$th write request revealed to the
        malicious servers is $W_i = \{w^{(j)}_i\}_{j \in \Adv_S}$.

        In the next steps of the game, the challenger will randomly reorder the honest
        clients' write requests. 
        The challenger should learn nothing about which client wrote what,
        despite all the information at its disposal.

        \item
        The challenger then samples a random permutation $\pi$ over
        $\{0, \dots, |\mathcal{H}_c|-1 \}$.
        The challenger sends the following set of write requests to the adversary,
        permuted according to~$\pi$:
        \[ \langle W_{\pi(0)}, W_{\pi(1)}, \dots, W_{\pi(|\HH_c|-1)} \rangle \]
        \end{itemize}

  \item For each client $i$ in $\{0, \dots, n-1\} \setminus \HH_c$, 
        the adversary computes a write
        request $(w^{(0)}_i, \dots, w^{(s-1)}_{i})$ (possibly according 
        to some malicious strategy) and sends the set of these
        write requests to the challenger.
  
  \item \begin{itemize}
        \item For each server $j \in \HH_s$, the
        challenger computes the server's final state
        $\sigma_j$ by running the ${\sf Update}$
        functionality on each of the $n$ client write requests in order.
        Let $S = \{ (j, \sigma_j) \ |\ j \in \HH_s \}$ be the set of 
        states of the honest servers.

        \item 
        The challenger samples a bit $b \rgets \{0,1\}$. 
        If $b = 0$, the challenger send $(S, \pi)$ to the adversary.
        Otherwise, the challenger samples a fresh permutation
        $\pi^*$ on $\HH_c$ and sets $(S, \pi^*)$ to the adversary.

        \end{itemize}
  \item The adversary makes a guess $b'$ for the value of $b$.
\end{enumerate}

The adversary wins the game if $b = b'$.
We define the adversary's advantage as
$|\Pr[b = b'] - 1/2|$.
The scheme maintains $(s,t)$-write privacy if no efficient adversary
wins the game with non-negligible advantage (in the implicit security parameter).

 \section{Correctness Proof for $(2,1)$-DPF}
\label{app:correct}

This appendix proves correctness of the distributed
point construction of Section~\ref{sec:dpf:twoserver}.
For the scheme to be correct,
it must be that, for
$(k_A, k_B) \gets {\sf Gen}(\ell, m)$, for all $\ell' \in \Z_L$:
\[{\sf Eval}(k_A, \ell') + {\sf Eval}(k_B, \ell') = P_{\ell, m}(\ell'). \]
Let $(\ell_x,\ell_y)$ be the tuple in $\Z_x \times \Z_y$ representing
location $\ell$ and let $(\ell'_x,\ell'_y)$ be the tuple representing $\ell'$.
Let:
\begin{align*}
m'_A \gets {\sf Eval}(k_A, \ell') \qquad m'_B \gets {\sf Eval}(k_B, \ell').
\end{align*}
We use a case analysis to show that the left-hand side of the 
equation above equals $P_{\ell,m}$ for all $\ell'$:
\begin{itemize}
\item[{\bf Case I}:] $\ell_x \neq \ell'_x$. 
  When $\ell_x \neq \ell'_x$, 
  the seeds ${\bf s}_A[\ell'_x]$ and ${\bf s}_B[\ell'_x]$
  are equal, so ${\bf g}_A = {\bf g}_B$.
  Similarly ${\bf b}_A[\ell'_x] = {\bf b}_B[\ell'_x]$.
  The output $m'_A$ will be
  ${\bf g}_A[\ell'_y] + {\bf b}_A[\ell'_x]{\bf v}[\ell'_y]$,
  The output $m'_B$ will be identical to $m'_A$.
  Since the field is a binary field, adding
  a value to itself results in the zero element,
  so the sum $m'_A + m'_B$ will be zero as desired.

\item[{\bf Case II}:] $\ell_x = \ell'_x$ and $\ell_y \neq \ell'_y$.
  When $\ell_x = \ell'_x$, the seeds ${\bf s}_A[\ell'_x]$ and ${\bf s}_B[\ell'_x]$
  are {\em not} equal, so ${\bf g}_A \neq {\bf g}_B$.
  Similarly ${\bf b}_A[\ell'_x] \neq {\bf b}_B[\ell'_x]$.
  When $\ell_y \neq \ell'_y$, 
  ${\bf v}[\ell'_y] = {\bf g}_A[\ell'_y] + {\bf g}_B[\ell'_y]$.
  Assume ${\bf b}_A[\ell'_x] = 0$ (an analogous argument applies when 
  ${\bf b}_A[\ell'_x] = 1$), then:
  \[  {\bf v}[\ell'_y] = (m \cdot e_{\ell_y})[\ell'_y] + {\bf g}_A[\ell'_y] + {\bf g}_B[\ell'_y]. \]
  The sum $m'_A + m'_B$ will then be:
  \[ m'_A + m'_B = {\bf g}_A[\ell'_y] + {\bf g}_B[\ell'_y] + {\bf v}[\ell'_y] = 0.\]

\item[{\bf Case III}:] $\ell_x = \ell'_x$ and $\ell_y = \ell'_y$.
  This is the same as Case II, except that $(m \cdot e_{\ell_y})[\ell'_y] = m$ when
  $\ell_y = \ell'_y$, so the sum $m'_A + m'_B = m$, as desired.

\end{itemize}

\section{Proof Sketches for the\\{\sf AlmostEqual} Protocol}
\label{app:almostequal}

This appendix proves security of the ${\sf AlmostEqual}$ protocol 
of Section~\ref{sec:disrupt:smc}.

\nicepara{Completeness}
We must show that 
if the vectors ${\bf v}_A$ and ${\bf v}_B$ differ
in exactly one position, the audit server will output ``1''
with overwhelming probability.

The audit server checks three things:
(1) that $\mA$ and $\mB$ differ at a single index, 
(2) that the check values $c_A$ and $c_B$ are equal, and
(3) that the digests $d_A$ and $d_B$ match the digests
of the vectors the database servers sent to the audit server.
The second and third tests will always pass if all parties are honest.

The first test fails with some small probability:
since the audit server only outputs ``1'' if {\em exactly}
one element of the test vectors is equal, if there
is a collision in the hash function, the
protocol may return an incorrect result.
The probability of this event is negligible
in the security parameter $\lambda$.

\nicepara{Soundness}
To demonstrate soundness, it is sufficient to show that it is infeasible
for a cheating client to produce values
$(\hvA, \hvB, \hsA, \hsB, \hat{d}_A, \hat{d}_B)$
such that $\hvA$ and $\hvB$ do {\em not} differ at exactly one index, and
yet the three honest servers accept these vectors as almost equal.

Let $(\hmA, \hcA)$ and $(\hmB, \hcB)$ 
be the values that the honest database servers send 
to the audit server.

First, note that for the cheating client to succeed, it must be that
$\hat{d}_A = H(\hmA)$ or $\hat{d}_B = H(\hmB)$.
So the choice of $\hat{d}_A$ and $\hat{d}_B$ are fixed by the client's choice
of the other values.

Next, observe that if a cheating client submits values $\hsA \neq \hsB$ to
the servers, then $\hcA \neq \hcB$ with probability $1$.
This holds because we require soundness to hold only when all three
servers are honest, so both database servers will generate
$c_A$ and $c_B$ using a common blinding value $\rho$.
So any successfully cheating client must submit a request with
$\hcA = \hcB$.

To cause the servers to accept, the client must then find values 
$(\hvA, \hsA)$ and $(\hvB, \hsB)$ such that,
$\sigma = \hsA = \hsB$ and, for all but one ${i \in \{0, \dots, n-1\}}$, 
$\hmA[i] = \hmB[i]$.
If $\hvA = \hvB$ everywhere, then the probability of this event is zero:
the vectors $\hmA$ and $\hmB$ will always be the same and the servers will
always reject.

Consider instead that $\hvA$ and $\hvB$ differ at two or more indices
and yet the servers accept the vectors.
In this case, $\hvA$ and $\hvB$ differ at least at indices
$i_1$ and $i_2$ and yet $\hmA$ and $\hmB$ differ at only one index.
In this case, at some $i^* \in \{i_1, i_2\}$, we have that
\[ \hvA[i^*] \neq \hvB[i^*]\quad \text{ and }\quad H(\hvA[i^*], \hat{r}_{i^*}) = H(\hvB[i^*], \hat{r}_{i^*}).\]
If the client can find 
such vectors $\hvA$ and $\hvB$, then the client can find a collision
in $H$.
Any adversary that violates the soundness property of
the protocol with non-negligible probability can then 
find collisions in $H$ with non-negligible probability,
which is infeasible.

\nicepara{Zero Knowledge}
To show the zero-knowledge property we must show
that 
(1) each database server can simulate its view
of the protocol run and 
(2) the audit server can simulate
its view as well.
Each simulator takes as input all public parameters of
the system plus each server's private input.
Each simulator must produce faithful simulations whenever
the other parties are honest.

\medskip

Showing that the audit server can simulate its view
of the protocol run is straightforward: the audit
server receives values $(\mA, c_A, d_A)$ and $(\mB, c_B, d_B)$
from the database servers and honest client.

Whenever the vectors differ at exactly one position the audit
server can simulate its view of the protocol.
To do so, the simulator picks length-$n$
vectors $\mA$ and $\mB$ of random elements in 
the range of the hash function $H$
subject to the constraint
that the vectors are equal everywhere except 
at a random index $i' \in \Z_n$.
The simulator outputs the two vectors as
the vectors received from servers $A$ and $B$.
The simulator computes the digests $d_A = H({\bf m}_A)$
and $d_B = H({\bf m}_B)$.
The simulator sets $c_A = c_B \rgets \lbits$.

The simulation of $\mA$ and $\mB$ is perfect, since 
these values are independently random values, as long as
all of the values $(r_0, \dots, r_{n-1})$ generated from
the seed $\sigma$ are distinct.
Since the servers sample each $r_i$ from $\lbits$, the
probability of a collision is negligible.
The digests $d_A$ and $d_B$ are constructed exactly as in the 
real protocol run.
The values $c_A$ and $c_B$ that the auditor sees
in the real protocol are equal values masked by a
random $\lambda$-bit string.
The simulation of $c_A$ and $c_B$ is then perfect.

\medskip

We now must show that each database server can simulate
its view as well.
Since the role of the two database servers is symmetric, we need only
produce a simulator for server $A$.

In the real protocol run, the database server interacts
with the honest client and honest audit server as follows:
\begin{enumerate}
  \item The honest client sends $\sigma$ to the database server.
  \item The database server produces a (possibly malformed) test vector
    $\hmA$ and check value $\hat{c}_A$, and sends these values to the audit server.
  \item The audit server returns a bit $\hbet$ to the database server.
\end{enumerate}

We must produce a simulator $S = (S_1, S_2)$ that simulates the first
and third flows of this protocol.
The simulator takes all of the public parameters of the system as implicit
input.
The simulator also takes the vector $\vA$ and the shared random value
$\rho$ as input.
The simulation proceeds as follows:
\begin{itemize}
  \item $(\sigma, \state) \gets S_1(\vA, \rho)$. 
        Simulator $S_1$ produces a random PRG seed
        $\sigma \in \lbits$. 
        The simulator computes $\mA$ using $\vA$ and $\sigma$, 
        as in the real protocol.
        The simulator computes $c_A \gets \rho \oplus \sigma$
        and $d_A \gets H(\mA)$.
        The simulator outputs $\vA$ as the output of the first flow, and outputs
        the state as:
        ${\state = (d_A, c_A)}$.
      \item $\hat{\beta} \gets S_2(\state = (d_A, c_A), \hmA, \hat{c}_A)$.
        The simulator computes ${\hat{d}_A = H(\hmA)}$ and outputs
        $\hat{\beta} = 1$ if 
        \[ d_A = \hat{d}_A\quad\text{ and }\quad \hat{c}_A = c_A.\]
        The simulator outputs
        $\hat{\beta} = 0$ otherwise.
\end{itemize}

We now must argue that the simulation is accurate.
The simulation of $\sigma$ is perfect, since
this is just a random $\lambda$-bit value in the real protocol.

The distribution of $\hat{\beta}$ is 
statistically close to the distribution in the real protocol.
Whenever $c_A \neq \hat{c}_A$, in the real protocol, the audit
server will return $0$, as in the simulation.
Whenever $d_A \neq H(\hmA)$, in the real protocol, the audit
server will return $0$, as in the simulation.
So the only time when the simulation and real protocol diverge
is when $d_A = H(\hmA)$ but $\hmA \neq \mA$.
The probability (over the randomness of server $A$ and 
the random oracle) that $A$ outputs such a vector $\hmA$
after making a polynomial number of random-oracle queries
is negligible in the security parameter.
Thus, the simulation is near perfect.

 \abbr{}{
\section{Security Proof Sketches for the\\Three-Server Protocol}
\label{app:smc-proof}

This appendix contains the security proofs for
the three-server protocol for detecting malicious
client requests (Section~\ref{sec:disrupt:smc}).

\nicepara{Completeness}
If the pair of keys is well-formed then 
the ${\bf b}_A$ and ${\bf b}_B$ vectors
(also the ${\bf s}_A$ and ${\bf s}_B$ vectors)
are equal at every index $i \neq \ell_x$ and they 
differ at index $i = \ell_x$.
Even in the negligibly unlikely event that the random seed
chosen at ${\bf s}_A[\ell_x]$ is equal to the random seed 
chosen at ${\bf s}_B[\ell_x]$, the test vectors ${\bf t}_A$ and ${\bf t}_B$
will still differ because ${\bf b}_A[\ell_x] \neq {\bf b}_B[\ell_x]$.
Thus, a correct pair of ${\bf b}$ and ${\bf s}$ vectors will pass the first 
${\sf AlmostEqual}$ check.

The second ${\sf AlmostEqual}$ check is more subtle.
If the ${\bf v}$ vector is well formed then, 
letting $\ell_x$ be the index where
the ${\bf s}$ vectors differ, we have:
\begin{align*}
  {\bf u}_B &= \left(\sum_{i=0}^{x-1} G({\bf s}_A[i]) \right) +   
                  G({\bf s}_A[\ell_x]) + G({\bf s}_B[\ell_x]) + {\bf v}\\
      &= {\bf u}_A + G({\bf s}_A[\ell_x]) + G({\bf s}_B[\ell_x]) + {\bf v}\\
&= {\bf u}_A + m \cdot e_{\ell_y}
\end{align*}
If ${\bf v}$ is well-formed, 
then two test vectors ${\bf u}_A$ and ${\bf u}_B$
differ only at index $\ell_y$.

\nicepara{Soundness}
To show soundness, we must bound the probability that the audit
server will output ``1'' when the servers take a malformed pair 
of DPF keys as input.
If the ${\bf b}$ and ${\bf s}$ vectors are not equal everywhere
except at one index, the soundness of the ${\sf AlmostEqual}$ protocol
implies that the audit server will return ``0'' with overwhelming
probability when invoked the first time.

Now, given that the ${\bf s}$ vectors differ at one index, we
can demonstrate that if the ${\bf u}$ vectors pass the second
${\sf AlmostEqual}$ check, then ${\bf v}$ is also well formed.
Let $\ell_x$ be the index at which the ${\bf s}$ vectors differ.
Write the values of the ${\bf s}$ vectors at index $\ell_x$
as $s^*_A$ and $s^*_B$.
Then, by construction:
\begin{align*}
  {\bf u}_A &= \left(\sum_{i \neq \ell_x}^{x-1} G({\bf s}_A[i])\right) + G(s^*_A)\\
  {\bf u}_B &= \left(\sum_{i \neq \ell_x}^{x-1} G({\bf s}_B[i])\right) + G(s^*_B) + {\bf v}
\end{align*}
The first term of these two expressions are equal (because the
${\bf s}$ vectors are equal almost everywhere).
Thus, to violate the soundness property, an adversary must construct
a tuple $(s^*_A, s^*_B, {\bf v})$ such that the vectors
$G(s^*_A)$ and $(G(s^*_B) + {\bf v})$ differ at exactly one
index {\em and} such that ${\bf v} \neq G(s^*_A) + G(s^*_B) + m \cdot e_{\ell}$.
This is a contradiction, however, since if 
$G(s^*_A)$ and $(G(s^*_B) + {\bf v})$ differ at exactly one index,
then: 
\begin{align*}
  m \cdot e_{\ell_y} = G(s^*_A) + [(G(s^*_B) + {\bf v})]
\end{align*}
for some $\ell_y$ and $m$, by definition of $m \cdot e_{\ell_y}$.

\nicepara{Zero Knowledge}
The audit server can simulate its view of a successful
run of the protocol (one in which the input keys are
well-formed) by invoking the ${\sf AlmostEqual}$ simulator twice.
 \section{Security Proof Sketches for the\\Zero-Knowledge Protocol}
\label{app:zkp-proof}

\nicepara{Completeness}
Completeness for the first half of the protocol,
which checks the form of the ${\bf B}$ and ${\bf S}$ vectors,
follows directly from the construction of those vectors.

The one slightly subtle step comes in Step~\ref{item:check-zkp}
of the second half of the protocol. 
For the protocol to be complete, it must be that
${\bf G}_\textrm{sum}$ is zero at every index except one.
This is true because:
\begin{align*}
{\bf G}_\textrm{sum} = (\Sigma_{i=0}^{s-1} {\bf G}_i) + {\bf v} &= G(s^*) + m \cdot e_{\ell_y} - G(s^*) = m \cdot e_{\ell_y}
\end{align*}

\nicepara{Soundness}
The soundness of the non-interactive zero-knowledge proof 
in the first half of the protocol guarantees that
the ${\bf B}$ vectors sum to $e_{\ell_x}$ and that
the ${\bf s}$ vectors sum to $s^* \cdot e_{\ell_x}$ for
some values $\ell_x \in \Z_x$ and $s^* \in \Seed$.

We must now argue that the probability that all servers
accept an invalid write request in the second half of the 
protocol is negligible.
The soundness property of the underlying
zero-knowledge proof used in Step~\ref{item:check-zkp}
implies that the vector ${\bf G}_\textrm{sum}$ contains
commitments to zero at all indices except one.
A client who violates the soundness property produces
a vector ${\bf v}$ and seed value $s^*$ such
that $(\Sigma_{i=0}^{s-1} {\bf G}_i) + {\bf v} = m \cdot e_{\ell_y}$
for some values $\ell_y \in \Z_y$ and $m\in\G$, and that
${\bf v} \neq m \cdot e_{\ell_y} - G(s^*)$.
This is a contradiction, however, since 
$(\Sigma_{i=0}^{s-1} {\bf G}_i) = G(s^*)$, by the first
half of the protocol, and so:
\begin{align*}
(\Sigma_{i=0}^{s-1} {\bf G}_i) + {\bf v} = m \cdot e_{\ell_y} = G(s^*) + {\bf v}
\end{align*}
Finally, we conclude that ${\bf v} = m \cdot e_{\ell_y} - G(s^*)$.

\nicepara{Zero Knowledge}
The servers can simulate every message they receive during
a run of the protocol.
In particular, they see only Pedersen commitments, which
are statistically hiding, and non-interactive zero-knowledge
proofs, which are simulatable in the random-oracle 
model~\cite{bellare1993random}.

 }

\end{document}